\documentclass[prd,nofootinbib,english,a4paper]{revtex4-2}
\usepackage[utf8]{inputenc}
\usepackage[T1]{fontenc}
\usepackage{amsmath}
\usepackage{amsfonts}
\usepackage{amssymb}
\usepackage{amsthm}
\usepackage{accents}
\usepackage{enumitem}
\usepackage[colorlinks=true]{hyperref}
\usepackage{tikz}

\allowdisplaybreaks

\newcommand{\lc}[1]{\accentset{\circ}{#1}}
\newcommand{\dd}{\mathrm{d}}

\begin{document}

\title{Homogeneous and isotropic cosmology in general teleparallel gravity}

\author{Lavinia Heisenberg}
\email{lavinia.heisenberg@phys.ethz.ch}
\affiliation{Institut f\"{u}r Theoretische Physik, Philosophenweg 16, 69120 Heidelberg, Germany}
\affiliation{Institute for Theoretical Physics, ETH Zurich, Wolfgang-Pauli-Strasse 27, 8093 Zurich, Switzerland}
\affiliation{Perimeter Institute, 31 Caroline Street N, Waterloo ON, Canada}

\author{Manuel Hohmann}
\email{manuel.hohmann@ut.ee}
\affiliation{Laboratory of Theoretical Physics, Institute of Physics, University of Tartu, W. Ostwaldi 1, 50411 Tartu, Estonia}

\author{Simon Kuhn}
\email{simkuhn@phys.ethz.ch}
\affiliation{Institute for Theoretical Physics, ETH Zurich, Wolfgang-Pauli-Strasse 27, 8093 Zurich, Switzerland}
\affiliation{Perimeter Institute, 31 Caroline Street N, Waterloo ON, Canada}

\begin{abstract}
We derive the most general homogeneous and isotropic teleparallel geometries, defined by a metric and a flat, affine connection. We find that there are five branches of connection solutions, which are connected via several limits, and can further be restricted to the torsion-free and metric-compatible cases. We apply our results to several classes of general teleparallel gravity theories and derive their cosmological dynamics for all five branches. Our results show that for large subclasses of these theories the dynamics reduce to that of closely related metric or symmetric teleparallel gravity theories, while for other subclasses up to two new scalar degrees of freedom participate in the cosmological dynamics.
\end{abstract}

\maketitle
%\tableofcontents

\section{Motivation}\label{sec:intro}
Various cosmological observations, such as the tension between late-time and early-time measurements of the Hubble parameter~\cite{Planck:2018vyg}, show indications towards physics beyond the so-called $\Lambda$CDM model, which describes the dynamics of the universe through general relativity, a cosmological constant $\Lambda$ and cold dark matter (CDM), and has become widely accepted as the standard model of cosmology. In order to explain these observations, besides introducing new and unknown types of matter, numerous modified gravity theories have been developed and their cosmological dynamics has been studied~\cite{DiValentino:2021izs,Saridakis:2021lqd,Heisenberg:2018vsk}. Most conventionally, these theories depart from the common formulation of general relativity using the metric and its Levi-Civita connection, thereby attributing gravity to the curvature of the latter. However, besides these also a large number of so-called teleparallel theories exists, which attribute gravity to the torsion or nonmetricity of a flat connection, and these three possibilities have been subsumed under the title ``geometric trinity of gravity''~\cite{BeltranJimenez:2019tjy}. Further, instead of restricting the teleparallel geometry to being either metric-compatible or symmetric, one may also consider geometries which allow for both torsion and nonmetricity, leading to the class of general teleparallel gravity theories~\cite{BeltranJimenez:2019odq,Boehmer:2021aji}. Several classes of such theories have been proposed and studied~\cite{BeltranJimenez:2021auj,BeltranJimenez:2021kpj,Hohmann:2022mlc}.

In order to study the cosmological dynamics of any given gravity theory, one usually assumes that all dynamical fields present in the theory exhibit cosmological symmetry, i.e., homogeneity and isotropy, and are thus invariant under spatial rotations and translations. For the metric, this leads to the well-known Friedmann-Lemaître-Robertson-Walker form, while, e.g., a scalar field is assumed to be constant along hypersurfaces of constant time. For teleparallel gravity theories, one of the dynamical fields is a flat, affine connection, and so one must also impose homogeneity and isotropy on this field for consistency. Previous works in this direction have shown that the most general homogeneous and isotropic affine connection is described by five functions of time, two of which can be associated with torsion, while three are related to nonmetricity~\cite{Minkevich:1998cv,Hohmann:2019fvf}. If one imposes vanishing curvature and torsion, one finds three branches of solutions described by one function of time~\cite{Hohmann:2021ast,DAmbrosio:2021pnd}, while for vanishing curvature and nonmetricity one finds three branches without any additional free functions besides those determining the metric~\cite{Hohmann:2020zre}.

The aim of this article is to continue the aforementioned line of studies and determine the most general homogeneous and isotropic teleparallel geometry, defined by a metric and a flat, affine connection, where the latter is restricted only by the condition of vanishing curvature, but may possess both torsion and nonmetricity, and to study the physical consequences of our findings in the context of different general teleparallel gravity theories. Moreover, we study how a coupling between matter and the teleparallel connection, which gives rise to a hypermomentum described as a cosmological hyperfluid~\cite{Iosifidis:2020gth}, enters as a source into the cosmological field equations.

The outline of this article is as follows. In section~\ref{sec:general} we give a brief review of general teleparallel gravity, its dynamical fields and the general structure of the field equations. Bianchi identities are discussed in section~\ref{sec:bianchi}. We then construct the most general class of homogeneous and isotropic geometries in section~\ref{sec:cosmo}. In section~\ref{eq:matter}, we apply our findings to the matter side of the field equations and study the possible energy-momentum-hypermomentum sources. Further, in section~\ref{sec:examples} we discuss several classes of general teleparallel gravity theories, derive their cosmological dynamics and study their general properties as well as relate them to previously studied classes of theories. Finally, we give an example for a possible contribution of the newly introduced connection degrees of freedom to the cosmological dynamics in section~\ref{sec:solution}. We end with a conclusion in section~\ref{sec:conclusion}.

\section{General teleparallel gravity}\label{sec:general}
We start by giving a brief review of the class of general teleparallel gravity theories. In their metric-affine formulation, the dynamical fields are given by a Lorentzian metric \(g_{\mu\nu}\) and an affine connection with coefficients \(\Gamma^{\mu}{}_{\nu\rho}\), which is imposed to be flat,
\begin{equation}\label{eq:nocurv}
R^{\rho}{}_{\sigma\mu\nu} = \partial_{\mu}\Gamma^{\rho}{}_{\sigma\nu} - \partial_{\nu}\Gamma^{\rho}{}_{\sigma\mu} + \Gamma^{\rho}{}_{\lambda\mu}\Gamma^{\lambda}{}_{\sigma\nu} - \Gamma^{\rho}{}_{\lambda\nu}\Gamma^{\lambda}{}_{\sigma\mu} = 0\,.
\end{equation}
Note that this connection is different from the Levi-Civita connection, whose coefficients are the Christoffel symbols
\begin{equation}
\lc{\Gamma}^{\mu}{}_{\nu\rho} = \frac{1}{2}g^{\mu\sigma}(\partial_{\nu}g_{\sigma\rho} + \partial_{\rho}g_{\nu\sigma} - \partial_{\sigma}g_{\nu\rho})\,,
\end{equation}
where we use an empty circle to denote any quantity related to the Levi-Civita connection. Their difference can be written as
\begin{equation}
\Gamma^{\mu}{}_{\nu\rho} - \lc{\Gamma}^{\mu}{}_{\nu\rho} = M^{\mu}{}_{\nu\rho} = K^{\mu}{}_{\nu\rho} + L^{\mu}{}_{\nu\rho}\,,
\end{equation}
where the distortion \(M^{\mu}{}_{\nu\rho}\) is composed of the contortion
\begin{equation}\label{eq:contortion}
K^{\mu}{}_{\nu\rho} = \frac{1}{2}\left(T_{\nu}{}^{\mu}{}_{\rho} + T_{\rho}{}^{\mu}{}_{\nu} - T^{\mu}{}_{\nu\rho}\right)\,,
\end{equation}
as well as the disformation
\begin{equation}\label{eq:disformation}
L^{\mu}{}_{\nu\rho} = \frac{1}{2}\left(Q^{\mu}{}_{\nu\rho} - Q_{\nu}{}^{\mu}{}_{\rho} - Q_{\rho}{}^{\mu}{}_{\nu}\right)\,,
\end{equation}
and these are defined through the torsion
\begin{equation}\label{eq:torsion}
T^{\mu}{}_{\nu\rho} = \Gamma^{\mu}{}_{\rho\nu} - \Gamma^{\mu}{}_{\nu\rho}\,,
\end{equation}
and the nonmetricity
\begin{equation}\label{eq:nonmetricity}
Q_{\mu\nu\rho} = \nabla_{\mu}g_{\nu\rho} = \partial_{\mu}g_{\nu\rho} - \Gamma^{\sigma}{}_{\nu\mu}g_{\sigma\rho} - \Gamma^{\sigma}{}_{\rho\mu}g_{\nu\sigma}\,.
\end{equation}
The dynamical fields, among which we also include an arbitrary set \(\psi^I\) of matter fields, constitute the action, which we write in the form
\begin{equation}
S[g, \Gamma, \psi] = S_{\text{g}}[g, \Gamma] + S_{\text{m}}[g, \Gamma, \psi]\,,
\end{equation}
where \(S_{\text{g}}\) is the gravitational action, which defines the gravity theory under consideration, and \(S_{\text{m}}\) is a generic matter action. It follows that the variation of the latter can be written in the form
\begin{equation}\label{eq:matactvar}
\delta S_{\text{m}} = \int_M\left(\frac{1}{2}\Theta^{\mu\nu}\delta g_{\mu\nu} + H_{\mu}{}^{\nu\rho}\delta\Gamma^{\mu}{}_{\nu\rho} + \Psi_I\delta\psi^I\right)\sqrt{-g}\dd^4x\,,
\end{equation}
where \(\Psi_I = 0\) are the matter field equations, and we introduced the energy-momentum tensor \(\Theta_{\mu\nu}\) and the hypermomentum \(H_{\mu}{}^{\nu\rho}\). For the gravitational part of the action, we similarly write
\begin{equation}\label{eq:metricgravactvar}
\delta S_{\text{g}} = -\int_M\left(\frac{1}{2}W^{\mu\nu}\delta g_{\mu\nu} + Y_{\mu}{}^{\nu\rho}\delta\Gamma^{\mu}{}_{\nu\rho}\right)\sqrt{-g}\dd^4x\,,
\end{equation}
where \(W^{\mu\nu}\) and \(Y_{\mu}{}^{\nu\rho}\) depend on the gravity theory under consideration. When performing the variation with respect to the connection, one must take into account its flatness~\eqref{eq:nocurv}, either by adding a Lagrange multiplier and keeping the variation \(\delta\Gamma^{\mu}{}_{\nu\rho}\) of the connection arbitrary, or by allowing only variations of the form
\begin{equation}\label{eq:metaffflatvar}
\delta\Gamma^{\mu}{}_{\nu\rho} = \nabla_{\rho}\xi^{\mu}{}_{\nu}
\end{equation}
with a tensor field \(\xi^{\mu}{}_{\nu}\), which preserve the flatness by construction. Following either approach yields the connection field equations~\cite{Hohmann:2021fpr,Hohmann:2022mlc}
\begin{equation}\label{eq:genconnfield}
\nabla_{\tau}Y_{\mu}{}^{\nu\tau} - M^{\omega}{}_{\tau\omega}Y_{\mu}{}^{\nu\tau} = \nabla_{\tau}H_{\mu}{}^{\nu\tau} - M^{\omega}{}_{\tau\omega}H_{\mu}{}^{\nu\tau}\,,
\end{equation}
while variation with respect to the metric yields the field equations
\begin{equation}\label{eq:genfield}
W_{\mu\nu} = \Theta_{\mu\nu}\,.
\end{equation}
As mentioned before, the dependence of the terms \(W_{\mu\nu}\) and \(Y_{\mu}{}^{\nu\rho}\) on the dynamical fields depends on the choice of the gravitational action, and thus the theory under consideration. A simple example is the general teleparallel equivalent of general relativity (GTEGR), whose action reads \cite{BeltranJimenez:2019odq}
\begin{equation}
S_{\text{g}} = -\frac{1}{2\kappa^2}\int\dd^4x\sqrt{-g}G\,,
\end{equation}
where the scalar \(G\) in the action is defined as
\begin{multline}\label{eq:gtegr}
G = 2M^\mu{}_{\nu[\mu}M^{\nu\rho}{}_{\rho]} = \bigg(\frac{1}{4}Q^{\mu\nu\rho}Q_{\mu\nu\rho} - \frac{1}{2}Q^{\mu\nu\rho}Q_{\rho\mu\nu} - \frac{1}{4}Q^{\rho\mu}{}_{\mu}Q_{\rho\nu}{}^{\nu} + \frac{1}{2}Q^{\mu}{}_{\mu\rho}Q^{\rho\nu}{}_{\nu}\\
+ \frac{1}{4}T^{\mu\nu\rho}T_{\mu\nu\rho} + \frac{1}{2}T^{\mu\nu\rho}T_{\rho\nu\mu} - T^{\mu}{}_{\rho\mu}T_{\nu}{}^{\rho\nu} + T^{\mu\nu\rho}Q_{\nu\rho\mu} - T^{\mu}{}_{\rho\mu}Q_{\rho\nu}{}^{\nu} + T^{\mu}{}_{\rho\mu}Q^{\nu}{}_{\nu\rho}\bigg)\,.
\end{multline}
This action has the property that it agrees with the Einstein-Hilbert action up to a boundary term \(B\),
\begin{equation}
\lc{R} = -G + B\,, \quad
B = 2\lc{\nabla}_{\mu}M^{[\nu\mu]}{}_{\nu} = -\lc{\nabla}_{\mu}(Q^{\mu\nu}{}_{\nu} - Q_{\nu}{}^{\nu\mu} - 2T_{\nu}{}^{\nu\mu})\,,
\end{equation}
and so its field equations turn out to be identical to those of general relativity~\cite{Hohmann:2022mlc},
\begin{equation}
W_{\mu\nu} = \frac{1}{\kappa^2}\left(\lc{R}_{\mu\nu} - \frac{1}{2}\lc{R}g_{\mu\nu}\right)\,, \quad
\nabla_{\tau}Y_{\mu}{}^{\nu\tau} - M^{\omega}{}_{\tau\omega}Y_{\mu}{}^{\nu\tau} = 0\,.
\end{equation}
We will discuss a number of generalizations of GTEGR in section~\ref{sec:examples}.

\section{Bianchi identity of general teleparallel theories}\label{sec:bianchi}
In GR the famous Bianchi identity states that the Einstein tensor is automatically covariantly conserved, and so is the energy-momentum tensor of matter by the Einstein equation. Here we want to derive the Bianchi identity of a general teleparallel theory and look at the consequences for hypermomentum conservation.

Let us assume again we have an action $S[g,\Gamma,\psi]$ of metric, connection, and matter fields. This action can be either $S_g$, $S_m$, or their sum, where the first is of course independent of $\psi$. In each case the action is by construction invariant under coordinate transformations, i.e. its value will not change if we perform an infinitesimal coordinate transformation $x^\mu\to x^\mu+\zeta^\mu$. But the integrand in the action will change by its Lie derivative along $\zeta^\mu$, so we obtain
\begin{equation}
    0=\delta_\zeta S[g,\Gamma,\psi]=\int_M\left(-\frac12\mathcal{E}^{\mu\nu}\mathcal{L}_\zeta g_{\mu\nu}+\mathcal{Y}_\alpha{}^{\mu\nu}\mathcal{L}_\zeta\Gamma^\alpha{}_{\mu\nu}+\mathcal{P}_I\mathcal{L}_\zeta\psi^I\right)\sqrt{-g}\dd^4x\ .
\end{equation}
Let us also define
\begin{equation}
    \mathcal{C}_\mu{}^\nu=\nabla_\tau \mathcal{Y}_\mu{}^{\nu\tau}-M^\omega{}_{\tau\omega}\mathcal{Y}_\mu{}^{\nu\tau}\ .
\end{equation}
If we take for $S$ the full action then $\mathcal{E}_{\mu\nu}=0$, $\mathcal{C}_\mu{}^\nu=0$, and $\mathcal{P}_I=0$ will be the equations of motion of metric, connection, and matter fields, respectively. For either of the three choices of $S$ we will have $\mathcal{P}_I=0$ either trivially or by the matter field equation of motion $\Psi_I=0$, so we can drop this term; otherwise we must consider concrete field theories to find the form of $\mathcal{L}_\zeta\psi^I$. For the other two Lie derivatives we have
\begin{align}
    \mathcal{L}_\zeta g_{\mu\nu}&=\zeta^\lambda\partial_\lambda g_{\mu\nu}+2\partial_{(\mu}\zeta^\lambda g_{\nu)\lambda}=2\lc\nabla_{(\mu}\zeta_{\nu)}\,,\\
    \mathcal{L}_\zeta\Gamma^\alpha{}_{\mu\nu}&=\zeta^\lambda\partial_\lambda\Gamma^\alpha{}_{\mu\nu}+\partial_\mu\zeta^\lambda\Gamma^\alpha{}_{\lambda\nu}+\partial_\nu\zeta^\lambda\Gamma^\alpha{}_{\mu\lambda}-\partial_\lambda\zeta^\alpha\Gamma^\lambda{}_{\mu\nu}+\partial_\mu\partial_\nu\zeta^\alpha=R^\alpha{}_{\mu\lambda\nu}\zeta^\lambda+\nabla_\nu\nabla_\mu\zeta^\alpha+\nabla_\nu(T^\alpha{}_{\lambda\mu}\zeta^\lambda)\ .
\end{align}
For our flat connection the Riemann tensor vanishes, so we have
\begin{align}
    0&=\delta_\zeta S[g,\Gamma,\psi]=\int_M\left(-\mathcal{E}^{\mu\nu}\lc\nabla_\mu\zeta_\nu+\mathcal{Y}_\alpha{}^{\mu\nu}\nabla_\nu(\nabla_\mu\zeta^\alpha+T^\alpha{}_{\lambda\mu}\zeta^\lambda)\right)\sqrt{-g}\dd^4x=\\
    &=\int_M\left(\lc\nabla_\mu\mathcal{E}^\mu{}_\nu+(\nabla_\mu-M^\omega{}_{\mu\omega})\mathcal{C}_\nu{}^\mu-T^\alpha{}_{\nu\mu}\mathcal{C}_\alpha{}^\mu\right)\zeta^\nu\,\sqrt{-g}\dd^4x\ ,
\end{align}
where we integrated by parts\footnote{Note that by the identity
\begin{equation}
    \partial_\mu(\sqrt{-g}\,V^\mu)=\sqrt{-g}(\lc\nabla_\mu V^\mu)=\sqrt{-g}(\nabla_\mu V^\mu-M^\mu{}_{\nu\mu}V^\nu)\nonumber
\end{equation}
for any vector $V^\mu$, integrating the connection $\nabla$ by parts will introduce additional terms proportional to $M^\mu{}_{\nu\mu}$. We also assume here that boundary terms vanish, e.g., that $\zeta^\nu$ vanishes on the boundary}. Since this must hold for any $\zeta^\mu$ we obtain the Bianchi identity
\begin{equation}\label{eq:Bianchi}
    \lc\nabla_\mu\mathcal{E}^\mu{}_\nu+(\nabla_\mu-M^\omega{}_{\mu\omega})\mathcal{C}_\nu{}^\mu-T^\alpha{}_{\nu\mu}\mathcal{C}_\alpha{}^\mu=0\,,
\end{equation}
or in other form
\begin{equation}
    \lc\nabla_\mu(\mathcal{E}^\mu{}_\nu+\mathcal{C}_\nu{}^\mu)+(M^\alpha{}_{\nu\mu}-T^\alpha{}_{\nu\mu})\mathcal{C}_\alpha{}^\mu=0\,.
\end{equation}
This identity trivially holds if we take for $S$ the full action $S_g+S_m$, although one finds that if the metric equation alone is fulfilled we still have
\begin{equation}
    \nabla_\mu\mathcal{C}_\nu{}^\mu-T^\alpha{}_{\nu\mu}\mathcal{C}_\alpha{}^\mu=0
\end{equation}
automatically. In theories without torsion the connection equation of motion is actually just $(\nabla_\mu-M^\omega{}_{\mu\omega})\mathcal{C}_\nu{}^\mu=0$, which is by the Bianchi identity fulfilled automatically whenever the metric equations are. Hence they become obsolete by the Bianchi identity. If we consider only $S=S_g$ we have $\mathcal{E}^{\mu\nu}=W^{\mu\nu}$ and $\mathcal{Y}_\alpha{}^{\mu\nu}=Y_\alpha{}^{\mu\nu}$, and~\eqref{eq:Bianchi} becomes a non-trivial identity between the left side of the equations of motion. One can check that it holds for all examples given below. When using the matter action $S_m$ we have $\mathcal{E}^{\mu\nu}=\Theta^{\mu\nu}$, and we assume the matter energy-momentum tensor to be covariantly conserved on its own, $\lc\nabla_\mu\Theta^\mu{}_\nu=0$. We then find that hypermomentum must be covariantly conserved as well in the form
\begin{equation}
    \nabla_\mu\mathcal{C}_\nu{}^\mu-T^\alpha{}_{\nu\mu}\mathcal{C}_\alpha{}^\mu=0\ , \mathcal{C}_\mu{}^\nu=\nabla_\tau H_\mu{}^{\nu\tau}-M^\omega{}_{\tau\omega} H_\mu{}^{\nu\tau}\,.
\end{equation}
Lastly we consider the case when the gravitational action also depends on a scalar field $\Phi$, as we will discuss below. Then $\mathcal{L}_\zeta\Phi=\zeta^\mu\partial_\mu\Phi=\zeta^\mu\lc\nabla\Phi$, and we find the Bianchi identity
\begin{equation}
    \lc\nabla_\mu\mathcal{E}^\mu{}_\nu+\nabla_\mu\mathcal{C}_\nu{}^\mu-T^\alpha{}_{\nu\mu}\mathcal{C}_\alpha{}^\mu+\Psi\lc\nabla_\nu\Phi=0\,,
\end{equation}
where $\Psi$ is the equation of motion of $\Phi$.

%\textbf{Should we add the hypermomentum conservation equations to the examples below? Remove this if unnecessary.}

\section{Homogeneous and isotropic cosmology}\label{sec:cosmo}
In this section we derive the most general teleparallel geometry which is compatible with the assumption of cosmological symmetry, following closely the approach used for the metric and symmetric teleparallel cases~\cite{Hohmann:2020zre,Hohmann:2021ast,DAmbrosio:2021pnd}. Using spherical coordinates \((t,r,\vartheta,\varphi)\), the generating vector fields establishing cosmological symmetry are the three rotation generators
\begin{subequations}\label{eq:genrot}
\begin{align}
\varrho_1 &= \sin\varphi\partial_{\vartheta} + \frac{\cos\varphi}{\tan\vartheta}\partial_{\varphi}\,,\\
\varrho_2 &= -\cos\varphi\partial_{\vartheta} + \frac{\sin\varphi}{\tan\vartheta}\partial_{\varphi}\,,\\
\varrho_3 &= -\partial_{\varphi}\,,
\end{align}
\end{subequations}
as well as the three translation generators
\begin{subequations}\label{eq:gentra}
\begin{align}
\tau_1 &= \chi\sin\vartheta\cos\varphi\partial_r + \frac{\chi}{r}\cos\vartheta\cos\varphi\partial_{\vartheta} - \frac{\chi\sin\varphi}{r\sin\vartheta}\partial_{\varphi}\,,\\
\tau_2 &= \chi\sin\vartheta\sin\varphi\partial_r + \frac{\chi}{r}\cos\vartheta\sin\varphi\partial_{\vartheta} + \frac{\chi\cos\varphi}{r\sin\vartheta}\partial_{\varphi}\,,\\
\tau_3 &= \chi\cos\vartheta\partial_r - \frac{\chi}{r}\sin\vartheta\partial_{\vartheta}\,.
\end{align}
\end{subequations}
We then demand that both the metric \(g_{\mu\nu}\) and the flat affine connection \(\Gamma^{\mu}{}_{\nu\rho}\), which constitute the dynamical fields in the teleparallel geometry, are invariant under the action of these vector fields. This means that their Lie derivatives~\cite{Yano:1957lda}
\begin{equation}\label{eq:metsymcondi}
(\mathcal{L}_{X}g)_{\mu\nu} = X^{\rho}\partial_{\rho}g_{\mu\nu} + \partial_{\mu}X^{\rho}g_{\rho\nu} + \partial_{\nu}X^{\rho}g_{\mu\rho}
\end{equation}
and
\begin{equation}\label{eq:affsymcondi}
(\mathcal{L}_{X}\Gamma)^{\mu}{}_{\nu\rho} = X^{\sigma}\partial_{\sigma}\Gamma^{\mu}{}_{\nu\rho} - \partial_{\sigma}X^{\mu}\Gamma^{\sigma}{}_{\nu\rho} + \partial_{\nu}X^{\sigma}\Gamma^{\mu}{}_{\sigma\rho} + \partial_{\rho}X^{\sigma}\Gamma^{\mu}{}_{\nu\sigma} + \partial_{\nu}\partial_{\rho}X^{\mu}\\
\end{equation}
must vanish, where we used the abbreviation \(\chi = \sqrt{1 - u^2r^2}\), and \(u^2 \in \mathbb{R}\) indicates the spatial curvature. The latter can take positive or negative values, so that \(u\) can be real or imaginary; this choice of the parameter will turn out to be more practical than the more common curvature parameter \(k = u^2\). It is well known that the most general metric which satisfies the condition~\eqref{eq:metsymcondi} is the Robertson-Walker metric, whose non-vanishing components are given by
\begin{equation}\label{eq:metcosmo}
g_{tt} = -N^2\,, \quad
g_{rr} = \frac{A^2}{\chi^2}\,, \quad
g_{\vartheta\vartheta} = A^2r^2\,, \quad
g_{\varphi\varphi} = g_{\vartheta\vartheta}\sin^2\vartheta\,,
\end{equation}
where we denote by \(N = N(t)\) the lapse function and by \(A = A(t)\) the scale factor. Note that we will keep the former general at this point, and make a convenient choice later in this article. Further, we introduce the hypersurface conormal \(n_{\mu}\) and spatial metric \(h_{\mu\nu}\), which allow us to decompose the metric in the form
\begin{equation}\label{eq:metricsplit}
g_{\mu\nu} = -n_{\mu}n_{\nu} + h_{\mu\nu}\,.
\end{equation}
Their non-vanishing components are given by
\begin{equation}
n_t = -N\,, \quad
h_{rr} = \frac{A^2}{\chi^2}\,, \quad
h_{\vartheta\vartheta} = A^2r^2\,, \quad
h_{\varphi\varphi} = h_{\vartheta\vartheta}\sin^2\vartheta\,.
\end{equation}
Further, the Levi-Civita tensor \(\epsilon_{\mu\nu\rho\sigma}\) of \(g_{\mu\nu}\) gives rise to spatial Levi-Civita tensor \(\varepsilon_{\mu\nu\rho}\) via
\begin{equation}
\varepsilon_{\mu\nu\rho} = n^{\sigma}\epsilon_{\sigma\mu\nu\rho}\,, \quad
\epsilon_{\mu\nu\rho\sigma} = 4\varepsilon_{[\mu\nu\rho}n_{\sigma]}\,.
\end{equation}
For the affine connection, we will proceed in two steps. In addition to the symmetry condition~\eqref{eq:affsymcondi} for the six generating vector fields of cosmological symmetry, we must also impose the condition~\eqref{eq:nocurv} of vanishing curvature. Starting with the former condition, one finds that the non-vanishing components of the most general cosmologically symmetric affine connection are given by~\cite{Minkevich:1998cv,Hohmann:2019fvf,DAmbrosio:2021pnd}
\begin{gather}
\Gamma^t{}_{tt} = K_1\,, \quad
\Gamma^{\vartheta}{}_{r\vartheta} = \Gamma^{\vartheta}{}_{\vartheta r} = \Gamma^{\varphi}{}_{r\varphi} = \Gamma^{\varphi}{}_{\varphi r} = \frac{1}{r}\,, \quad
\Gamma^{\varphi}{}_{\vartheta\varphi} = \Gamma^{\varphi}{}_{\varphi\vartheta} = \cot\vartheta\,, \quad
\Gamma^{\vartheta}{}_{\varphi\varphi} = -\sin\vartheta\cos\vartheta\,,\nonumber\\
\Gamma^t{}_{rr} = \frac{K_2}{\chi^2}\,, \quad
\Gamma^r{}_{\vartheta\vartheta} = -r\chi^2\,, \quad
\Gamma^r{}_{\varphi\varphi} = -r\chi^2\sin^2\vartheta\,, \quad
\Gamma^r{}_{\varphi\vartheta} = -\Gamma^r{}_{\vartheta\varphi} = K_5r^2\chi\sin\vartheta\,,\nonumber\\
\Gamma^t{}_{\vartheta\vartheta} = K_2r^2\,, \quad
\Gamma^r{}_{tr} = \Gamma^{\vartheta}{}_{t\vartheta} = \Gamma^{\varphi}{}_{t\varphi} = K_3\,, \quad
\Gamma^r{}_{rt} = \Gamma^{\vartheta}{}_{\vartheta t} = \Gamma^{\varphi}{}_{\varphi t} = K_4\,, \quad
\Gamma^r{}_{rr} = \frac{kr}{\chi^2}\,,\nonumber\\
\Gamma^t{}_{\varphi\varphi} = K_2r^2\sin^2\vartheta\,, \quad
\Gamma^{\vartheta}{}_{r\varphi} = -\Gamma^{\vartheta}{}_{\varphi r} = \frac{K_5\sin\vartheta}{\chi}\,, \quad
\Gamma^{\varphi}{}_{r\vartheta} = -\Gamma^{\varphi}{}_{\vartheta r} = -\frac{K_5}{\chi\sin\vartheta}\,,\label{eq:cosmoaffconn}
\end{gather}
where \(K_1(t), \ldots, K_5(t)\) are functions of time. For this connection, we can now calculate the curvature tensor, which reads
\begin{multline}
R^{\mu}{}_{\nu\rho\sigma} = 2\frac{K_3(K_4 - K_1) + \partial_tK_3}{N^2}n_{\nu}n_{[\rho}h^{\mu}_{\sigma]}
+ 2\frac{K_2(K_4 - K_1) - \partial_tK_2}{A^2}n^{\mu}n_{[\rho}h_{\sigma]\nu}\\
+ 2\frac{K_2K_5N}{A^3}n^{\mu}\varepsilon_{\nu\rho\sigma}
- 2\frac{K_3K_5}{NA}n_{\nu}\varepsilon^{\mu}{}_{\rho\sigma}
- 2\frac{\partial_tK_5}{NA}\varepsilon^{\mu}{}_{\nu[\rho}n_{\sigma]}
+ 2\frac{u^2 + K_2K_3 - K_5^2}{A^2}h^{\mu}_{[\rho}h_{\sigma]\nu}\,,
\end{multline}
making use of the decomposition~\eqref{eq:metricsplit} we introduced before. Note that all terms appearing in this decomposition are independent, and so their coefficients must vanish separately. Hence, the connection is flat if and only if
\begin{equation}
\partial_tK_5 = K_2K_5 = K_3K_5 = u^2 + K_2K_3 - K_5^2 = K_3(K_4 - K_1) + \partial_tK_3 = K_2(K_4 - K_1) - \partial_tK_2 = 0\,.
\end{equation}
In order to determine the most general solution to these conditions, we distinguish the following two cases:
\begin{enumerate}
\item
We start by assuming \(u = 0\), i.e., vanishing spatial curvature. In this case we have the condition \(K_2K_3 = K_5^2\), so either both sides are vanishing or non-vanishing. However, from \(K_2K_5 = K_3K_5 = 0\) follows that \(K_5 = 0\) or \(K_2 = K_3 = 0\). Hence, the only option is \(K_5 = K_2K_3 = 0\). Therefore, at least one of \(K_2\) or \(K_3\) must vanish, which leaves three cases to be distinguished:
\begin{enumerate}
\item
Setting \(K_2 = K_3 = 0\), the remaining equations are already satisfied and no further restrictions on the parameter functions arise. \(K_1\) and \(K_4\) are the only parameters left, which are arbitrary and unconstrained.
\item
For \(K_3 = 0\) and \(K_2 \neq 0\), \(K_2\) is a free function. Now only one of \(K_1\) and \(K_4\) is left undetermined, since the difference is constrained to satisfy
\begin{equation}
K_4 - K_1 = \frac{\partial_tK_2}{K_2}\,.
\end{equation}
\item
Assuming \(K_2 = 0\) and \(K_3 \neq 0\), one obtains a similar result as in the previous case, but now \(K_3\) is a free function and the difference between \(K_1\) and \(K_4\) must satisfy
\begin{equation}
K_4 - K_1 = -\frac{\partial_tK_3}{K_3}\,.
\end{equation}
\end{enumerate}
\item
We then continue with the spatially curved case \(u \neq 0\). Now we can further distinguish the following two cases:
\begin{enumerate}
\item
\(K_5 \neq 0\): From \(K_2K_5 = K_3K_5 = 0\) follows \(K_2 = K_3 = 0\). Hence, \(K_5 = \pm u\), which becomes imaginary for negative curvature $u^2<0$, and the remaining equations are satisfied. \(K_1\) and \(K_4\) are left undetermined.
\item
\(K_5 = 0\): In this case one has \(K_2K_3 = -u^2 \neq 0\) and so both must be non-zero and inversely proportional, so that only one of them can be chosen arbitrarily. This further implies
\begin{equation}
0 = \partial_tK_2K_3 + K_2\partial_tK_3\,,
\end{equation}
and so
\begin{equation}
K_4 - K_1 = \frac{\partial_tK_2}{K_2} = -\frac{\partial_tK_3}{K_3}\,,
\end{equation}
so that the remaining equations consistently determine the difference between \(K_1\) and \(K_4\).
\end{enumerate}
\end{enumerate}
We see that there are five different branches, in each of which the flat, cosmologically symmetric connection is determined by two functions of time. In the following, however, it will turn out to be more convenient to introduce a different parametrization, which is based on the expressions
\begin{equation}
T^{\mu}{}_{\nu\rho} = 2T_1h^{\mu}_{[\nu}n_{\rho]} + 2T_2\varepsilon^{\mu}{}_{\nu\rho}\,, \quad
Q_{\rho\mu\nu} = 2Q_1n_{\rho}n_{\mu}n_{\nu} + 2Q_2n_{\rho}h_{\mu\nu} + 2Q_3h_{\rho(\mu}n_{\nu)}\,.
\end{equation}
for the most general cosmologically symmetric torsion and nonmetricity tensors~\cite{Minkevich:1998cv,Iosifidis:2020gth}, where \(T_1, T_2, Q_1, Q_2, Q_3\) are functions of time \(t\), which serve as an alternative parametrization to the previously introduced \(K_1, \ldots, K_5\). It follows from the tensor character of the expressions on both sides of the equations that these new quantities are scalars under coordinate transformations, which makes them more suitable as dynamical variables in the cosmological field equations. They are related to the previously introduced quantities by
\begin{equation}
T_1 = \frac{K_4 - K_3}{N}\,, \quad
T_2 = \frac{K_5}{A}\,, \quad
Q_1 = \frac{\partial_tN}{N^2} - \frac{K_1}{N}\,, \quad
Q_2 = \frac{1}{N}\left(K_4 - \frac{\partial_tA}{A}\right)\,, \quad
Q_3 = \frac{K_3}{N} - \frac{K_2N}{A^2}\,.
\end{equation}
From the case distinction discussed above, we see that in all cases we are free to choose the function \(K_4\), or equivalently, \(Q_2\) in the scalar variables, so that we can use it as a dynamical variable in all branches, which we denote by \(K = Q_2\). For the remaining branches of flat connections, we need to introduce different parametrizations, in order to express the remaining scalar variables in terms of one additional, independent, dynamical variable. We follow the same distinction as before, and use the Hubble parameter
\begin{equation}
H = \frac{\partial_tA}{NA}
\end{equation}
for convenience.
\begin{enumerate}
\item\label{it:flat}
For \(u = 0\), we have \(K_5 = 0\), and therefore \(T_2 = 0\). We then distinguish further:
\begin{enumerate}[ref=\theenumi{}\alph*]
\item\label{it:coin}
In the branch \(K_2 = K_3 = 0\), we find \(Q_3 = 0\). We are free to choose \(K_1\), and hence can define the second variable as \(L = Q_1\). The remaining scalar quantity is determined as \(T_1 = H + K\).
\item\label{it:para}
For \(K_2 \neq 0\), one has \(K_2\) as a free function, and so can use \(L = Q_3\) as dynamical variable. \(K_1\) is fixed, and so are
\begin{equation}
T_1 = H + K\,, \quad Q_1 = -K + H + \frac{\partial_tL}{NL}\,.
\end{equation}
\item\label{it:conf}
For \(K_3 \neq 0\), one can freely choose \(K_3\), which again allows to define \(L = Q_3\) as dynamical variable. In this case the remaining scalars read
\begin{equation}
T_1 = H + K - L\,, \quad Q_1 = -K - H - \frac{\partial_tL}{NL}\,.
\end{equation}
\end{enumerate}
\item\label{it:curv}
For the spatially curved case \(u \neq 0\), we use the following two parametrizations:
\begin{enumerate}[ref=\theenumi{}\alph*]
\item\label{it:axi}
For \(K_5 \neq 0\), we choose the sign of \(u\) in the parametrization such that \(K_5 = u\), and are free to choose \(L = Q_1\). The other sign convention of $u$ will give the same result with the replacement $u\to -u$. This yields
\begin{equation}
T_1 = H + K\,, \quad T_2 = \frac{u}{A}\,, \quad Q_3 = 0\,.
\end{equation}
\item\label{it:vec}
Finally, for \(K_5 = 0\), we can choose $L=H+K-T_1$ such that
\begin{equation}
T_1 = H + K - L\,, \quad T_2 = 0\,, \quad Q_1 = -K - H - \frac{\partial_tL}{NL}\,, \quad Q_3 = L + \frac{u^2}{A^2L}\,.
\end{equation}
\end{enumerate}
\end{enumerate}
To further simplify the parametrizations, it turns out to be convenient to introduce rescaled quantities
\begin{equation}
\mathcal{H} = AH\,, \quad \mathcal{K} = AK\,, \quad \mathcal{L} = AL\,, \quad \mathcal{T}_i = AT_i\,, \quad \mathcal{Q}_i = AQ_i\,,
\end{equation}
where \(\mathcal{H}\) is the conformal Hubble parameter, as well as the conformal time derivative
\begin{equation}
f' = \frac{A}{N}\partial_tf
\end{equation}
for any function \(f\) of time. In terms of these quantities, the parametrizations simplify, and take the form summarized in table~\ref{tab:branches}.

\begin{table}[htb]
\renewcommand*{\arraystretch}{2.5}
\begin{tabular}{|l|c|c|c||c|c|c|c|c|}
\hline
& $u$ & $K_2$ & $K_3$ & $\mathcal{T}_1$ & $\mathcal{T}_2$ & $\mathcal{Q}_1$ & $\mathcal{Q}_2$ & $\mathcal{Q}_3$\\\hline\hline
\ref{it:coin} & $= 0$ & $= 0$ & $= 0$ & $\mathcal{H} + \mathcal{K}$ & $0$ & $\mathcal{L}$ & $\mathcal{K}$ & $0$\\
\ref{it:para} & $= 0$ & $\neq 0$ & $= 0$ & $\mathcal{H} + \mathcal{K}$ & $0$ & $-\mathcal{K} + \dfrac{\mathcal{L}'}{\mathcal{L}}$ & $\mathcal{K}$ & $\mathcal{L}$\\
\ref{it:conf} & $= 0$ & $= 0$ & $\neq 0$ & $\mathcal{H} + \mathcal{K} - \mathcal{L}$ & $0$ & $-\mathcal{K} - \dfrac{\mathcal{L}'}{\mathcal{L}}$ & $\mathcal{K}$ & $\mathcal{L}$\\
\ref{it:axi} & $\neq 0$ & $= 0$ & $= 0$ & $\mathcal{H} + \mathcal{K}$ & $u$ & $\mathcal{L}$ & $\mathcal{K}$ & $0$\\
\ref{it:vec} & $\neq 0$ & $\neq 0$ & $\neq 0$ & $\mathcal{H} + \mathcal{K} - \mathcal{L}$ & $0$ & $-\mathcal{K} - \dfrac{\mathcal{L}'}{\mathcal{L}}$ & $\mathcal{K}$ & $\mathcal{L} + \dfrac{u^2}{\mathcal{L}}$\\
\hline
\end{tabular}
\caption{Branches for flat connections with cosmological symmetry.}
\label{tab:branches}
\end{table}

\begin{figure}[htb]
\begin{tikzpicture}
\filldraw[fill=black!20!white] (0,0) -- (-1.73,1) -- (-1.73,-1) -- (0,0);
\filldraw[fill=black!20!white] (0,0) to +(0.71,0.71) arc (135:-135:1) to +(-0.71,0.71);
\draw (-1.15,0) node {\ref{it:vec}};
\draw (1.15,0) node {\ref{it:axi}};
\draw (0,0) node [circle,draw,fill=black!10!white] {\ref{it:coin}};
\draw (-1.73,1) node [circle,draw,fill=black!10!white] {\ref{it:conf}};
\draw (-1.73,-1) node [circle,draw,fill=black!10!white] {\ref{it:para}};

\filldraw[fill=black!20!white] (-5,0) -- (-6.73,1) -- (-6.73,-1) -- (-5,0);
\draw (-6.15,0) node {\ref{it:vec}};
\draw (-5,0) node [circle,draw,fill=black!10!white] {\ref{it:coin}};
\draw (-6.73,1) node [circle,draw,fill=black!10!white] {\ref{it:conf}};
\draw (-6.73,-1) node [circle,draw,fill=black!10!white] {\ref{it:para}};

\filldraw[fill=black!20!white] (7,0) to +(0.71,0.71) arc (135:-135:1) to +(-0.71,0.71);
\filldraw[fill=black!20!white] (7,0) to +(-0.71,0.71) arc (45:315:1) to +(0.71,0.71);
\draw (5.85,0) node {\ref{it:vec}};
\draw (8.15,0) node {\ref{it:axi}};
\draw (7,0) node [circle,draw,fill=black!10!white] {\ref{it:flat}};

\draw[->] (2.75,0) -- ++(1.5,0) node [above,midway] {$Q_{\mu\nu\rho} \to 0$};
\draw[->] (-2.75,0) -- ++(-1.5,0) node [above,midway] {$T^{\mu}{}_{\nu\rho} \to 0$};
\end{tikzpicture}
\caption{Schematic relation between different cosmologically symmetric branches.}
\label{fig:branches}
\end{figure}
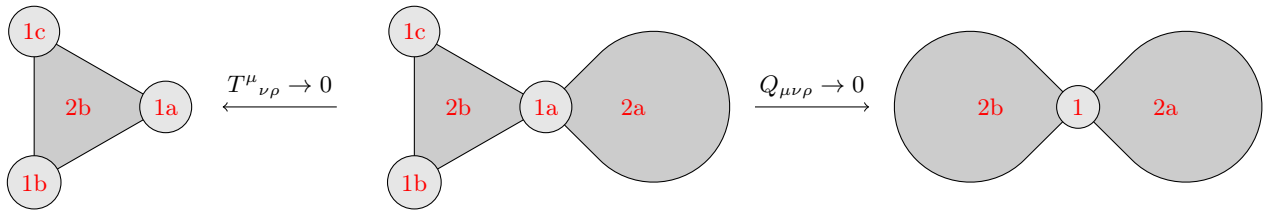

It is worth discussing how these different branches of cosmologies are related to each other. For the spatially curved branch~\ref{it:axi}, we see that \(K_2 = K_3 = 0\) everywhere, and so this property is preserved also in the limit \(u \to 0\), leading to the branch~\ref{it:coin}. This is different for the spatially curved branch~\ref{it:vec}. Using the parametrization given in table~\ref{tab:branches}, the limit \(u \to 0\) leads to the branch~\ref{it:conf}. However, this limit depends on the choice of the parametrization. By introducing a new parameter function \(\tilde{\mathcal{L}} = u^2/\mathcal{L}\), the parametrization of the branch~\ref{it:vec} becomes
\begin{equation}
\mathcal{T}_1 = \mathcal{H} + \mathcal{K} - \frac{u^2}{\tilde{\mathcal{L}}}\,, \quad
\mathcal{T}_2 = 0\,, \quad
\mathcal{Q}_1 = -\mathcal{K} + \frac{\tilde{\mathcal{L}}'}{\tilde{\mathcal{L}}}\,, \quad
\mathcal{Q}_2 = \mathcal{K}\,, \quad
\mathcal{Q}_3 = \tilde{\mathcal{L}} + \frac{u^2}{\tilde{\mathcal{L}}}\,.
\end{equation}
Taking the limit \(u \to 0\), and renaming \(\tilde{\mathcal{L}}\) to \(\mathcal{L}\), we obtain the branch~\ref{it:para}. Finally, defining a new parameter instead as \(\tilde{\mathcal{L}} = \mathcal{L}/u\), one obtains
\begin{equation}
\mathcal{T}_1 = \mathcal{H} + \mathcal{K} - u\tilde{\mathcal{L}}\,, \quad
\mathcal{T}_2 = 0\,, \quad
\mathcal{Q}_1 = -\mathcal{K} - \frac{\tilde{\mathcal{L}}'}{\tilde{\mathcal{L}}}\,, \quad
\mathcal{Q}_2 = \mathcal{K}\,, \quad
\mathcal{Q}_3 = u\left(\tilde{\mathcal{L}} + \frac{1}{\tilde{\mathcal{L}}}\right)\,.
\end{equation}
In the limit \(u \to 0\), we obtain the branch~\ref{it:coin}, with the identification of the new parameter
\begin{equation}
\mathcal{L} = -\mathcal{K} - \frac{\tilde{\mathcal{L}}'}{\tilde{\mathcal{L}}}\,.
\end{equation}
Further, one may consider the special cases of vanishing torsion and nonmetricity, respectively. First note that the branch~\ref{it:axi} has explicit torsion \(\mathcal{T}_2 = u\), and so does not have a limit with vanishing torsion. For the remaining branches, this limit is obtained by solving \(\mathcal{T}_1 = 0\) for \(\mathcal{K}\), and one obtains the four branches of symmetric teleparallel cosmology~\cite{Hohmann:2021ast,DAmbrosio:2021pnd}. Similarly, to obtain vanishing nonmetricity, one sets \(\mathcal{K} = 0\) and then solves for \(\mathcal{L}\). In this case the three spatially flat (\(u = 0\)) branches assume the common limit \(\mathcal{T}_1 = \mathcal{H}\) and \(\mathcal{T}_2 = 0\). Together with the two spatially curved branches, these agree with the result found for metric teleparallel cosmology~\cite{Hohmann:2020zre}. These relations are illustrated in figure~\ref{fig:branches}.

In the following sections, we will apply the conditions of homogeneity and isotropy to the matter source and gravitational field equations of teleparallel gravity theories.

\section{Cosmologically symmetric field equations and energy-momentum-hypermomentum}\label{eq:matter}
We apply the conditions of cosmological symmetry introduced in the previous section to the generic form~\eqref{eq:genfield} of the general teleparallel field equations. Assuming that the teleparallel geometry given by a metric and a flat affine connection is homogeneous and isotropic, it follows that also the left hand side of the gravitational field equations, which is constituted by these fields, has this symmetry, and must therefore be of the form
\begin{equation}
W_{\mu\nu} = \mathfrak{N}n_{\mu}n_{\nu} + \mathfrak{H}h_{\mu\nu}\,, \quad
%\nabla_{\tau}Y_{\mu\nu}{}^{\tau} - M^{\omega}{}_{\tau\omega}Y_{\mu\nu}{}^{\tau} = \mathfrak{T}n_{\mu}n_{\nu} + \mathfrak{S}h_{\mu\nu}\,,
\nabla_{\tau}Y_{\mu}{}^{\nu\tau} - M^{\omega}{}_{\tau\omega}Y_{\mu}{}^{\nu\tau} = \mathfrak{T}n_{\mu}n^{\nu} + \mathfrak{S}h_{\mu}{}^{\nu}\,,
\end{equation}
where the dependence of the scalars \(\mathfrak{N}, \mathfrak{H}, \mathfrak{T}, \mathfrak{S}\) follows from the gravity theory under consideration. For this form of hypermomentum the connection equation of motion becomes symmetric in $\mu$ and $\nu$. Hence, the right hand side of the field equations must be of the same form. For the energy-momentum tensor \(\Theta_{\mu\nu}\), the cosmological symmetry condition leads to the perfect fluid form
\begin{equation}
\Theta_{\mu\nu} = \rho n_{\mu}n_{\nu} + ph_{\mu\nu}\,.
\end{equation}
Similarly, imposing the cosmological symmetry on the hypermomentum leads to the form~\cite{Iosifidis:2020gth}
\begin{equation}
H_{\rho\mu\nu} = \phi h_{\mu\rho}n_{\nu} + \chi h_{\nu\rho}n_{\mu} + \psi h_{\mu\nu}n_{\rho} + \omega n_{\mu}n_{\nu}n_{\rho} - \zeta\varepsilon_{\rho\mu\nu}\,.
\end{equation}
Note that here and for the rest of the paper $\chi$ refers to the field appearing in the hypermomentum, and not to the expression $\sqrt{1-u^2r^2}$. Inserting these expressions into the gravitational field equations, one finds that they take the form
\begin{equation}
\mathfrak{N} = \rho\,, \quad
\mathfrak{H} = p
\end{equation}
for the metric equations, while the connection equations read
\begin{subequations}
\begin{align}
-A\mathfrak{T} &= \omega' + 3[\mathcal{H}\omega + (\mathcal{H} - \mathcal{T}_1 + \mathcal{Q}_2 - \mathcal{Q}_3)\psi + (\mathcal{H} - \mathcal{T}_1 + \mathcal{Q}_2)\chi]\,,\\
-A\mathfrak{S} &= \phi' + 3\mathcal{H}\phi + (\mathcal{H} - \mathcal{T}_1 + \mathcal{Q}_2 - \mathcal{Q}_3)\psi + (\mathcal{H} - \mathcal{T}_1 + \mathcal{Q}_2)\chi\,.
\end{align}
\end{subequations}
Note that the latter depend on the choice of the cosmologically symmetric connection. For the five different branches we found, they are given as follows. For the spatially curved branch~\ref{it:axi} and its flat limit~\ref{it:coin} we find
\begin{equation}
-A\mathfrak{T} = \omega' + 3\mathcal{H}\omega\,, \quad
-A\mathfrak{S} = \phi' + 3\mathcal{H}\phi\,.
\end{equation}
The spatially curved branch~\ref{it:vec} yields
\begin{equation}
-A\mathfrak{T} = \omega' + 3\mathcal{H}\omega + 3\mathcal{L}\chi - 3\frac{u^2}{\mathcal{L}}\psi\,, \quad
-A\mathfrak{S} = \phi' + 3\mathcal{H}\phi + \mathcal{L}\chi - \frac{u^2}{\mathcal{L}}\psi\,.
\end{equation}
For the spatially flat branch~\ref{it:conf} one has
\begin{equation}
-A\mathfrak{T} = \omega' + 3\mathcal{H}\omega + 3\mathcal{L}\chi\,, \quad
-A\mathfrak{S} = \phi' + 3\mathcal{H}\phi + \mathcal{L}\chi\,,
\end{equation}
while the branch~\ref{it:para} yields
\begin{equation}
-A\mathfrak{T} = \omega' + 3\mathcal{H}\omega - 3\mathcal{L}\psi\,, \quad
-A\mathfrak{S} = \phi' + 3\mathcal{H}\phi - \mathcal{L}\psi\,.
\end{equation}
We will make use of these expressions in the following sections, where we will derive the cosmological field equations for a number of example theories and discuss their physical implications.

\section{Application to teleparallel theories}\label{sec:examples}
We now apply our findings to a few classes of general teleparallel gravity theories and study their cosmological dynamics. In section~\ref{ssec:fg} we study the $f(G)$ class of theories, whose Lagrangian is given by an arbitrary function \(f\) of the GTEGR Lagrangian \(G\). Another class of theories based on the most general Lagrangian which is quadratic in torsion and nonmetricity is studied in section~\ref{ssec:gtqg}. Finally, we consider a class of theories with a non-minimally coupled scalar field in section~\ref{ssec:stg}.

\subsection{$f(G)$ gravity}\label{ssec:fg}
The first example we study is a generalization of GTEGR whose action is given by
\begin{equation}
S_{\text{g}} = -\frac{1}{2\kappa^2}\int\dd^4x\sqrt{-g}f(G)\,,
\end{equation}
where \(G\) denotes the scalar~\eqref{eq:gtegr}.
\begin{multline}
\delta G = \lc\nabla_\alpha\delta g^{\mu\nu}(M^\rho{}_{(\mu|\rho|}\delta^\alpha{}_{\nu)}-M^\alpha{}_{(\mu\nu)}+M^{[\alpha\rho]}{}_\rho g_{\mu\nu})+\delta\Gamma^\alpha{}_{\mu\nu}(-M^{\mu\nu}{}_\alpha-M^\nu{}_\alpha{}^\mu+M^\rho{}_{\alpha\rho}g^{\mu\nu}+M^{\mu\rho}{}_\rho\delta^\nu_\alpha)\\
+ \delta g^{\mu\nu}(M_{\rho(\mu\nu)}M^{\sigma\rho}{}_{\sigma} - M^{\rho}{}_{\sigma(\mu})M^{\sigma}{}_{\nu)\rho})\,.
\end{multline}
Varying the action with respect to the metric and the connection, one obtains the field equations~\cite{Hohmann:2022mlc}
\begin{subequations}
\begin{gather}
f'\left(\lc{R}_{\mu\nu} - \frac{1}{2}\lc{R}g_{\mu\nu}\right) - M^{\rho}{}_{(\mu\nu)}\lc{\nabla}_{\rho}f' + \lc{\nabla}_{(\mu}f'M^{\sigma}{}_{\nu)\sigma} + M^{[\rho\sigma]}{}_{\sigma}g_{\mu\nu}\lc{\nabla}_{\rho}f' + \frac{1}{2}(f - f'G)g_{\mu\nu} = \kappa^2\Theta_{\mu\nu}\,,\\
M^{\nu\rho}{}_{\rho}\nabla_{\mu}f' + M^{\sigma}{}_{\mu\sigma}g^{\nu\rho}\nabla_{\rho}f' - M^{\nu\rho}{}_{\mu}\nabla_{\rho}f' - M^{\rho}{}_{\mu}{}^{\nu}\nabla_{\rho}f' = 2\kappa^2(\nabla_{\rho}H_{\mu}{}^{\nu\rho} - M^{\omega}{}_{\rho\omega}H_{\mu}{}^{\nu\rho})\,.
\end{gather}
\end{subequations}
One can then evaluate these for the cosmologically symmetric branches of teleparallel geometries. First note that the scalar \(G\) takes the cosmological value
\begin{equation}
G = \frac{3}{A^2}[\mathcal{Q}_3(\mathcal{Q}_1 - \mathcal{Q}_2 + 2\mathcal{T}_1) + 2(\mathcal{Q}_2 - \mathcal{T}_1)^2 - 2\mathcal{T}_2^2]\,.
\end{equation}
The structure of the resulting cosmological field equations depends on the choice of the branch.

We start with the spatially curved branch~\ref{it:axi}, for which the metric field equations become
\begin{subequations}\label{eq:cosmofgaxi}
\begin{align}
12\mathcal{H}^2\frac{f'}{A^2} - f &= 2\kappa^2\rho\,,\\
-4\mathcal{E}\frac{f''}{A^4} - 4(\mathcal{H}' + 2\mathcal{H}^2 - u^2)\frac{f'}{A^2} + f &= 2\kappa^2p\,,
%-48\mathcal{H}^2(\mathcal{H}' - \mathcal{H}^2 + u^2)\frac{f''}{A^4} - 4(\mathcal{H}' + 2\mathcal{H}^2 - u^2)\frac{f'}{A^2} + f &= 2\kappa^2p\,,
\end{align}
\end{subequations}
with
\begin{equation}\label{eq:2aEGRelation}
\mathcal{E}=A^2\mathcal{H}G'\,,
\end{equation}
while the left hand side of the connection equations vanishes identically, so that they reduce to the continuity equations
\begin{equation}
\omega' + 3\mathcal{H}\omega = \phi' + 3\mathcal{H}\phi = 0\,.
\end{equation}
It is remarkable that the two functions \(\mathcal{K}\) and \(\mathcal{L}\) describing the dynamics of the flat connection do not explicitly appear in these equations. Calculating
\begin{equation}
G = \frac{6(\mathcal{H}^2 - u^2)}{A^2}\,,
\end{equation}
we see that these do not enter implicitly through \(f\) either, and so fully decouple for this branch. These findings also hold in the limit \(u \to 0\), in which the branch~\ref{it:coin} is obtained, for which the cosmological field equations become
\begin{subequations}\label{eq:cosmofgcoin}
\begin{align}
12\mathcal{H}^2\frac{f'}{A^2} - f &= 2\kappa^2\rho\,,\\
-4\mathcal{E}\frac{f''}{A^4} - 4(\mathcal{H}' + 2\mathcal{H}^2)\frac{f'}{A^2} + f &= 2\kappa^2p\,,
%-48\mathcal{H}^2(\mathcal{H}' - \mathcal{H}^2)\frac{f''}{A^4} - 4(\mathcal{H}' + 2\mathcal{H}^2)\frac{f'}{A^2} + f &= 2\kappa^2p\,,
\end{align}
\end{subequations}
and the scalar \(G\) becomes
\begin{equation}
G = 6\frac{\mathcal{H}^2}{A^2}\,,
\end{equation}
and~\eqref{eq:2aEGRelation} still holds. Note that these equations agree with those of the metric teleparallel class of \(f(T)\) theories; while the spatially curved case~\eqref{eq:cosmofgaxi} yield those for the ``axial'' branch of metric teleparallel cosmologies, the flat case~\eqref{eq:cosmofgcoin} yields its flat limit~\cite{Hohmann:2020zre,Bahamonde:2022ohm}.

The situation is similar, yet different for the remaining branches. For the spatially curved case~\ref{it:vec}, the metric cosmological field equations become
\begin{subequations}\label{eq:cosmofgvecm}
\begin{align}
-9(u^2 + \mathcal{L}^2)\mathcal{E}\frac{f''}{A^4\mathcal{L}^4} + \left[6\mathcal{H}\mathcal{L}(u^2 + 2\mathcal{H}\mathcal{L} - \mathcal{L}^2) - 3(u^2 + \mathcal{L}^2)\mathcal{L}'\right]\frac{f'}{A^2\mathcal{L}^2} - f &= 2\kappa^2\rho\,,\\
3(u^2 + 4\mathcal{H}\mathcal{L} - 3\mathcal{L}^2)\mathcal{E}\frac{f''}{A^4\mathcal{L}^4} + \left\{2\mathcal{L}[2\mathcal{L}(u^2 - 2\mathcal{H}^2) - 3\mathcal{H}(u^2 - \mathcal{L}^2)] - 4\mathcal{L}^2\mathcal{H}' + 3(u^2 + \mathcal{L}^2)\mathcal{L}'\right\}\frac{f'}{A^2\mathcal{L}^2} + f &= 2\kappa^2p\,,
\end{align}
\end{subequations}
while the connection equations become
\begin{subequations}\label{eq:cosmofgvecc}
\begin{align}
-9(u^2 + \mathcal{L}^2)\mathcal{E}\frac{f''}{A^4\mathcal{L}^4} &= \frac{2\kappa^2}{A}\left(\omega' + 3\mathcal{H}\omega + 3\mathcal{L}\chi - 3\frac{u^2}{\mathcal{L}}\psi\right)\,,\\
-3(u^2 + \mathcal{L}^2)\mathcal{E}\frac{f''}{A^4\mathcal{L}^4} &= \frac{2\kappa^2}{A}\left(\phi' + 3\mathcal{H}\phi + \mathcal{L}\chi - \frac{u^2}{\mathcal{L}}\psi\right)\,,
\end{align}
\end{subequations}
where
\begin{equation}
\mathcal{E} = 4\mathcal{H}\mathcal{L}^2(u^2 + \mathcal{H}\mathcal{L})(\mathcal{H} - \mathcal{L}) - 2\mathcal{L}^2(u^2 + 2\mathcal{H}\mathcal{L} - \mathcal{L}^2)\mathcal{H}' - 2u^2\mathcal{L}'^2 + \mathcal{L}(u^2 + \mathcal{L}^2)\mathcal{L}'' = -A^2\mathcal{L}^3\frac{G'}{3}\,,
\end{equation}
with
\begin{equation}
G = \frac{3}{\mathcal{L}^2A^2}[2\mathcal{L}(\mathcal{H} - \mathcal{L})(u^2 + \mathcal{H}\mathcal{L}) - (u^2 + \mathcal{L}^2)\mathcal{L}']\,.
\end{equation}
We now see that the scalar \(\mathcal{L}\) enters the field equations, while \(\mathcal{K}\) does not. Note that the connection field equations can be satisfied only if the hypermomentum satisfies the combined conservation equation
\begin{equation}
(\omega + 3\phi)' + 3\mathcal{H}(\omega + 3\phi) = 0\,.
\end{equation}
This holds also for the flat limiting cases; for the case~\ref{it:conf} we find the equations
\begin{subequations}\label{eq:cosmofgconf}
\begin{align}
-9\mathcal{L}\mathcal{E}\frac{f''}{A^4} + 3(4\mathcal{H}^2 - 2\mathcal{H}\mathcal{L} - \mathcal{L}')\frac{f'}{A^2} - f &= 2\kappa^2\rho\,,\\
3(4\mathcal{H} - 3\mathcal{L})\mathcal{E}\frac{f''}{A^4} - (4\mathcal{H}' - 3\mathcal{L}' - 6\mathcal{H}\mathcal{L} + 8\mathcal{H}^2)\frac{f'}{A^2} + f &= 2\kappa^2p\\
-9\mathcal{L}\mathcal{E}\frac{f''}{A^4} &= \frac{2\kappa^2}{A}(\omega' + 3\mathcal{H}\omega + 3\mathcal{L}\chi)\,,\\
-3\mathcal{L}\mathcal{E}\frac{f''}{A^4} &= \frac{2\kappa^2}{A}(\phi' + 3\mathcal{H}\phi + \mathcal{L}\chi)\,,
\end{align}
\end{subequations}
with
\begin{equation}
\mathcal{E} = 4\mathcal{H}^2(\mathcal{H} - \mathcal{L}) - 2(2\mathcal{H} - \mathcal{L})\mathcal{H}' + \mathcal{L}'' = -A^2\frac{G'}{3}\,,
\end{equation}
and
\begin{equation}
G = \frac{6\mathcal{H}(\mathcal{H} - \mathcal{L}) - 3\mathcal{L}'}{A^2}\,,
\end{equation}
while for the branch~\ref{it:para} we have
\begin{subequations}\label{eq:cosmofgpara}
\begin{align}
-9\mathcal{L}\mathcal{E}\frac{f''}{A^4} + 3(4\mathcal{H}^2 + 2\mathcal{H}\mathcal{L} + \mathcal{L}')\frac{f'}{A^2} - f &= 2\kappa^2\rho\,,\\
3(4\mathcal{H} + \mathcal{L})\mathcal{E}\frac{f''}{A^4} - (4\mathcal{H}' + 3\mathcal{L}' + 6\mathcal{H}\mathcal{L} + 8\mathcal{H}^2)\frac{f'}{A^2} + f &= 2\kappa^2p\\
-9\mathcal{L}\mathcal{E}\frac{f''}{A^4} &= \frac{2\kappa^2}{A}(\omega' + 3\mathcal{H}\omega - 3\mathcal{L}\psi)\,,\\
-3\mathcal{L}\mathcal{E}\frac{f''}{A^4} &= \frac{2\kappa^2}{A}(\phi' + 3\mathcal{H}\phi - \mathcal{L}\psi)\,,
\end{align}
\end{subequations}
with
\begin{equation}
\mathcal{E} = 4\mathcal{H}^2(\mathcal{H} + \mathcal{L}) - 2(2\mathcal{H} + \mathcal{L})\mathcal{H}' - \mathcal{L}'' = -A^2\frac{G'}{3}\,,
\end{equation}
and
\begin{equation}
G = \frac{6\mathcal{H}(\mathcal{H} + \mathcal{L}) + 3\mathcal{L}'}{A^2}\,,
\end{equation}
Note that the last three branches of cosmological dynamics allow for a particular classification of their solutions in case the right hand side of the connection field equations vanishes, e.g., for vanishing hypermomentum. Further assuming \(u^2 < 0\), i.e., negative spatial curvature, the connection field equations~\eqref{eq:cosmofgvecc} are solved by \(\mathcal{L} = \pm iu\). Choosing the lower sign, the metric field equations~\eqref{eq:cosmofgvecm} then reduce to
\begin{subequations}
\begin{align}
12\mathcal{H}(\mathcal{H} + iu)\frac{f'}{A^2} - f &= 2\kappa^2\rho\,,\\
-48(\mathcal{H} + iu)^2(\mathcal{H}' - \mathcal{H}^2 - iu\mathcal{H})\frac{f''}{A^4} - 4(\mathcal{H}' + 2\mathcal{H}^2 + 3iu\mathcal{H} - u^2)\frac{f'}{A^2} + f &= 2\kappa^2p\,,
\end{align}
\end{subequations}
while for the upper sign one obtains an equivalent set of equations up to the trivial redefinition \(u \mapsto -u\). It follows that in these cases the cosmological dynamics in \(f(G)\) gravity become identical to the vector branch of the metric teleparallel \(f(T)\) class of theories~\cite{Hohmann:2020zre,Bahamonde:2022ohm}. Since in this case the connection degree of freedom $\mathcal{L}$ is fixed by an algebraic equation we do not obtain a dynamical contribution of it to the metric evolution. So, even if the resulting equations would give rise to a dark energy like effect, i.e. accelerated expansion at late times, it should not be called dynamical as it would rather come from the choice of $f$ and not a new degree of freedom. Similarly, setting \(\mathcal{L} = 0\) solves the connection equations in~\eqref{eq:cosmofgconf} and~\eqref{eq:cosmofgpara} (again assuming that their right hand side vanishes), while in both cases the metric field equations reduce to~\eqref{eq:cosmofgcoin} again, and thus reproduce the flat limiting case of \(f(T)\) cosmology. Any remaining solutions, given by \(\mathcal{L} \neq \pm iu\) in the spatially curved case and \(\mathcal{L} \neq 0\) in the spatially flat cases - except in~\ref{it:coin} and~\ref{it:axi} where the connection equations of motion are trivial - must then satisfy \(\mathcal{E} = 0\), leading to $G=G_0$ where $G_0$ is a constant. This follows directly from the form of the connection equations of motion, where all terms on the left side are proportional to $f''\partial_\rho G$. In general the left side is then an algebraic combination of derivatives of $G$, but since in our cosmological setup $G$ depends only time all terms are proportional to $G'$. Since $G$ contains first derivatives of $\mathcal{L}$ the equation $G=G_0$ can be seen as a first order differential equation for $\mathcal{L}$. For example, for the case~\ref{it:para} one finds
\begin{equation}
    \mathcal{L}=\frac{\mathcal{L}_0}{A^2}+\frac{1}{3A^2}\int\dd \eta\,A^2(G_0A^2-6\mathcal{H}^2)\,,
\end{equation}
with $\mathcal{L}_0$ an integration constant and $\eta$ conformal time. However, one can easily see that for constant $G=G_0$ the metric equations of motion become
\begin{equation}
    \left(\lc R_{\mu\nu}-\frac12\lc R g_{\mu\nu}\right)=\frac{\kappa^2}{f'(G_0)}\Theta_{\mu\nu}+\frac{f(G_0)-f'(G_0)G_0}{2f'(G_0)}g_{\mu\nu}\ ,
\end{equation}
which are nothing but the GR Friedmann equations with rescaled $\kappa^2$ and an effective cosmological constant. This case does also not lead to a dynamical dark energy, since only $f$ and its derivatives evaluated at $G_0$ enter the metric equations, which are fixed and do not evolve, though one may see $G_0$ as another integration constant of solving $\mathcal{E}=0$ for $\mathcal{L}$. In the presence of hypermomentum the connection equations become more involved and will generally lead to $G'\neq0$ and more interesting solutions, but it is curious to see that in its absence $f(G)$ gravity does not involve a dynmical connection. Either $\mathcal{L}$ is fixed by an algebraic equation or it leads to constant $G$.

\subsection{General teleparallel quadratic gravity}\label{ssec:gtqg}
The next class of teleparallel gravity theories we study is known as general teleparallel quadratic gravity~\cite{BeltranJimenez:2019odq}. Its action is an arbitrary linear combination of the 11 scalars which are quadratic in torsion and nonmetricity, and hence takes the form
\begin{equation}
\begin{split}
S_{\text{g}} &= -\frac{1}{2\kappa^2}\int\dd^4x\sqrt{-g}\tilde{G}\\
&= -\frac{1}{2\kappa^2}\int\dd^4x\sqrt{-g}(c_1Q^{\mu\nu\rho}Q_{\mu\nu\rho} + c_2Q^{\mu\nu\rho}Q_{\rho\mu\nu} + c_3Q^{\rho\mu}{}_{\mu}Q_{\rho\nu}{}^{\nu} + c_4Q^{\mu}{}_{\mu\rho}Q_{\nu}{}^{\nu\rho} + c_5Q^{\mu}{}_{\mu\rho}Q^{\rho\nu}{}_{\nu}\\
&\phantom{=}+ a_1T^{\mu\nu\rho}T_{\mu\nu\rho} + a_2T^{\mu\nu\rho}T_{\rho\nu\mu} + a_3T^{\mu}{}_{\rho\mu}T_{\nu}{}^{\rho\nu} + b_1T^{\mu\nu\rho}Q_{\nu\rho\mu} + b_2T^{\mu\rho}{}_{\mu}Q_{\rho\nu}{}^{\nu} + b_3T^{\mu\rho}{}_{\mu}Q^{\nu}{}_{\nu\rho})\\
&= -\frac{1}{2\kappa^2}\int\dd^4x\sqrt{-g}[M^{\mu\nu\rho}(k_1M_{\mu\nu\rho} + k_2M_{\nu\rho\mu} + k_3M_{\mu\rho\nu} + k_4M_{\rho\nu\mu} + k_5M_{\nu\mu\rho})\\
&\phantom{=}+ k_6M_{\rho\mu}{}^{\mu}M^{\rho\nu}{}_{\nu} + k_7M_{\mu\rho}{}^{\mu}M^{\nu\rho}{}_{\nu} + k_8M^{\mu}{}_{\mu\rho}M_{\nu}{}^{\nu\rho} + k_9M_{\mu\rho}{}^{\mu}M_{\nu}{}^{\nu\rho} + k_{10}M^{\mu}{}_{\mu\rho}M^{\rho\nu}{}_{\nu} + k_{11}M_{\rho\mu}{}^{\mu}M^{\nu\rho}{}_{\nu}]\,,
\end{split}
\end{equation}
with 11 constant coefficients either parametrized as \(a_{1,\ldots,3}, b_{1,\ldots,3}, c_{1,\ldots,5}\) or \(k_{1\,\ldots,11}\). While the former parametrization gives more insight into the split between torsion and nonmetricity, and will be more convenient for the purposes of this article, the latter turns out to be more convenient to express the field equations, which can be written in the form~\cite{Hohmann:2022mlc}
\begin{equation}\label{eq:gngrmetfield}
U_{\mu\nu} - \lc{\nabla}_{\rho}(V^{\rho}{}_{\mu\nu}) + \frac{1}{2}\tilde{G}g_{\mu\nu} = \kappa^2\Theta_{\mu\nu}\,,
\end{equation}
and
\begin{equation}\label{eq:gngrconnfield}
\nabla_{\tau}Z_{\mu}{}^{\nu\tau} - M^{\omega}{}_{\tau\omega}Z_{\mu}{}^{\nu\tau} = 2\kappa^2(\nabla_{\tau}H_{\mu}{}^{\nu\tau} - M^{\omega}{}_{\tau\omega}H_{\mu}{}^{\nu\tau})\,,
\end{equation}
where we introduced the abbreviations
\begin{subequations}\label{eq:gngrvarabbrev}
\begin{multline}
U^{\mu\nu} = k_1(M^{\mu\rho\sigma}M^{\nu}{}_{\rho\sigma} - M^{\rho\mu}{}_{\sigma}M_{\rho}{}^{\nu\sigma} - M_{\rho\sigma}{}^{\mu}M^{\rho\sigma\nu}) - k_2M_{\rho}{}^{\sigma(\mu}M_{\sigma}{}^{\nu)\rho} + k_3(M^{\mu\rho\sigma}M^{\nu}{}_{\sigma\rho} - 2M^{\rho\sigma(\mu}M_{\rho}{}^{\nu)}{}_\sigma)\\
- k_4M^{\rho\mu}{}_{\sigma}M^{\sigma\nu}{}_{\rho} - k_5M^{\rho\sigma\mu}M_{\sigma\rho}{}^{\nu} + k_6M^{\mu\rho}{}_{\rho}M^{\nu\sigma}{}_{\sigma} - k_7M^{\rho\mu}{}_{\rho}M^{\sigma\nu}{}_{\sigma}\\
- k_8M_{\rho}{}^{\rho\mu}M_{\sigma}{}^{\sigma\nu} - k_9M_{\rho}{}^{\rho(\mu}M_{\sigma}{}^{\nu)\sigma} - (2k_6M_{\rho\sigma}{}^{\sigma} + k_{11}M_{\sigma\rho}{}^{\sigma} + k_{10}M^{\sigma}{}_{\sigma\rho})M^{\rho(\mu\nu)}\,,
\end{multline}
as well as
\begin{multline}
V^{\rho\mu\nu} = -2k_6g^{\rho(\mu}M^{\nu)\sigma}{}_{\sigma} - k_{11}g^{\rho(\mu}M_{\sigma}{}^{\nu)\sigma} - k_{10}M_{\sigma}{}^{\sigma(\mu}g^{\nu)\rho} + (k_4 - k_5 - k_1 - k_3)M^{(\mu\nu)\rho} + (k_5 - k_4 - k_1 - k_3)M^{(\mu|\rho|\nu)}\\
+ (k_1 - k_2 + k_3 - k_4 - k_5)M^{\rho(\mu\nu)} + \frac{1}{2}g^{\mu\nu}\big[(2k_6 - k_{10} - k_{11})M^{\rho\sigma}{}_{\sigma} + (k_{11} - 2k_7 - k_9)M^{\sigma\rho}{}_{\sigma} + (k_{10} - 2k_8 - k_9)M_{\sigma}{}^{\sigma\rho}\big]
\end{multline}
and
\begin{multline}
Z_{\mu}{}^{\nu\rho} = 2k_1M_{\mu}{}^{\nu\rho} + k_2(M^{\nu\rho}{}_{\mu} + M^{\rho}{}_{\mu}{}^{\nu}) + 2k_3M_{\mu}{}^{\rho\nu} + 2k_4M^{\rho\nu}{}_{\mu} + 2k_5M^{\nu}{}_{\mu}{}^{\rho} + 2k_6M_{\mu\sigma}{}^{\sigma}g^{\nu\rho} + 2k_7M^{\sigma\nu}{}_{\sigma}\delta_{\mu}^{\rho}\\
+ 2k_8M_{\sigma}{}^{\sigma\rho}\delta_{\mu}^{\nu} + k_9(M_{\sigma}{}^{\rho\sigma}\delta_{\mu}^{\nu} + M_{\sigma}{}^{\sigma\nu}\delta_{\mu}^{\rho}) + k_{10}(M^{\rho\sigma}{}_{\sigma}\delta_{\mu}^{\nu} + M^{\sigma}{}_{\sigma\mu}g^{\nu\rho}) + k_{11}(M^{\nu\sigma}{}_{\sigma}\delta_{\mu}^{\rho} + M^{\sigma}{}_{\mu\sigma}g^{\nu\rho})\,.
\end{multline}
\end{subequations}
The two parametrizations are related by
\begin{gather}
k_1 = 2a_1 - b_1 + 2c_1\,, \quad
k_2 = -2a_2 + b_1 + 2c_2\,, \quad
k_9 = -2a_3 + 2b_2 - b_3 + 2c_5\,, \quad
k_4 = a_2 + c_2\,, \quad
k_5 = a_2 - b_1 + 2c_1\,,\nonumber\\
k_6 = c_4\,, \quad
k_7 = a_3 + b_3 + c_4\,, \quad
k_8 = a_3 - 2b_2 + 4c_3\,, \quad
k_3 = -2a_1 + b_1 + c_2\,, \quad
k_{10} = -b_3 + 2c_5\,, \quad
k_{11} = b_3 + 2c_4\,.
\end{gather}
Note that for the values
\begin{equation}
a_1 = \frac{1}{4}\,, \quad
a_2 = \frac{1}{2}\,, \quad
a_3 = -1\,, \quad
b_1 = 1\,, \quad
b_2 = -1\,, \quad
b_3 = 1\,, \quad
c_1 = \frac{1}{4}\,, \quad
c_2 = -\frac{1}{2}\,, \quad
c_3 = -\frac{1}{4}\,, \quad
c_4 = 0\,, \quad
c_5 = \frac{1}{2}\,,
\end{equation}
or equivalently,
\begin{equation}
k_{11} = -k_2 = 1\,, \quad k_1 = k_3 = k_4 = k_5 = k_6 = k_7 = k_8 = k_9 = k_{10} = 0\,,
\end{equation}
we obtain the GTEGR action~\eqref{eq:gtegr}. We now study the cosmological dynamics of this class of theories. While deriving the cosmological field equations, one finds that not all 11 coefficients enter these equations independently, but only a limited number of linear combinations appears. A suitable parametrization of these linear combinations is given by
\begin{gather}
z_1 = -\frac{2a_1 + a_2 + 3a_3}{2}\,, \quad
z_2 = \frac{3}{2}(3a_2 + a_3 - 2a_1)\,, \quad
z_4 = 2a_1 + a_2 + 3a_3 - 2b_1 - 6b_2 + 4c_1 + 12c_3\,, \quad
z_7 = 2c_3 + c_5\,,\nonumber\\
z_3 = c_2 - c_4 + c_5\,, \quad
z_5 = b_1 + 3b_2 - 4c_1 - 12c_3\,, \quad
z_6 = b_2 + b_3 - 4c_3 - 2c_5\,, \quad
z_8 = c_1 + c_2 + c_3 + c_4 + c_5\,.
\end{gather}
This particular parametrization is chosen here as it will turn out to simplify the cosmological field equations as much as possible, and that for GTEGR it is simply given by
\begin{equation}
z_1 = 1\,, \quad z_2 = z_3 = z_4 = z_5 = z_6 = z_7 = z_8 = 0\,.
\end{equation}
Hence, we can easily see which new terms arise from modifications of GTEGR and how they alter the dynamics. We now display the cosmological field equations, and give a few remarks on their general properties. We start with the branch~\ref{it:axi}, for which they read
\begin{subequations}\label{eq:cosmoqaxi}
\begin{align}
6(z_1 + z_2)u^2 + 6(z_1 - 2z_6 - 4z_7)\mathcal{H}^2 - 6(z_6 + 2z_7)\mathcal{H}' - 6z_6\mathcal{K}' - 8z_8\mathcal{L}' &\nonumber\\
- 4z_8\mathcal{L}^2 - 3z_4\mathcal{K}^2 - 6(z_4 + z_5 + 2z_6)\mathcal{H}\mathcal{K} - 2(3z_6 + 6z_7 + 8z_8)\mathcal{H}\mathcal{L} - 6z_6\mathcal{K}\mathcal{L} &= 2\kappa^2A^2\rho\,,\\
-2(z_1 + z_2)u^2 - 2(z_1 + 2z_4 + 2z_5)\mathcal{H}^2 - 2(2z_1 + z_4 + z_5)\mathcal{H}' + 2z_5\mathcal{K}' + 4z_7\mathcal{L}' &\nonumber\\
- 3z_4\mathcal{K}^2 - 4z_8\mathcal{L}^2 - 2(3z_4 + z_5)\mathcal{H}\mathcal{K} - 2(3z_6 + 2z_7)\mathcal{H}\mathcal{L} - 6z_6\mathcal{K}\mathcal{L} &= 2\kappa^2A^2p\,,\\
-6(z_6 + 2z_7)(2\mathcal{H}^2 + \mathcal{H}') - 6z_6\mathcal{K}' - 8z_8\mathcal{L}' - 4\mathcal{H}(3z_6\mathcal{K} + 4z_8\mathcal{L}) &= 2\kappa^2A(\omega' + 3\mathcal{H}\omega)\,,\\
-2(z_4 + z_5)(2\mathcal{H}^2 + \mathcal{H}') - 2z_4\mathcal{K}' - 2z_6\mathcal{L}' - 4\mathcal{H}(z_4\mathcal{K} + z_6\mathcal{L}) &= 2\kappa^2A(\phi' + 3\mathcal{H}\phi)\,.
\end{align}
\end{subequations}
Note that both \(\mathcal{K}\) and \(\mathcal{L}\) appear as dynamical quantities with first-order time derivatives, and that the dynamics does not depend on the parameter \(z_3\) obtained from the action parameters. The same holds true for the branch~\ref{it:coin}, which is obtained taking the limit \(u \to 0\), and leads to the field equations
\begin{subequations}
\begin{align}
6(z_1 - 2z_6 - 4z_7)\mathcal{H}^2 - 6(z_6 + 2z_7)\mathcal{H}' - 6z_6\mathcal{K}' - 8z_8\mathcal{L}' &\nonumber\\
- 4z_8\mathcal{L}^2 - 3z_4\mathcal{K}^2 - 6(z_4 + z_5 + 2z_6)\mathcal{H}\mathcal{K} - 2(3z_6 + 6z_7 + 8z_8)\mathcal{H}\mathcal{L} - 6z_6\mathcal{K}\mathcal{L} &= 2\kappa^2A^2\rho\,,\\
-2(z_1 + 2z_4 + 2z_5)\mathcal{H}^2 - 2(2z_1 + z_4 + z_5)\mathcal{H}' + 2z_5\mathcal{K}' + 4z_7\mathcal{L}' &\nonumber\\
- 3z_4\mathcal{K}^2 - 4z_8\mathcal{L}^2 - 2(3z_4 + z_5)\mathcal{H}\mathcal{K} - 2(3z_6 + 2z_7)\mathcal{H}\mathcal{L} - 6z_6\mathcal{K}\mathcal{L} &= 2\kappa^2A^2p\,,\\
-6(z_6 + 2z_7)(2\mathcal{H}^2 + \mathcal{H}') - 6z_6\mathcal{K}' - 8z_8\mathcal{L}' - 4\mathcal{H}(3z_6\mathcal{K} + 4z_8\mathcal{L}) &= 2\kappa^2A(\omega' + 3\mathcal{H}\omega)\,,\\
-2(z_4 + z_5)(2\mathcal{H}^2 + \mathcal{H}') - 2z_4\mathcal{K}' - 2z_6\mathcal{L}' - 4\mathcal{H}(z_4\mathcal{K} + z_6\mathcal{L}) &= 2\kappa^2A(\phi' + 3\mathcal{H}\phi)\,.
\end{align}
\end{subequations}
Note that one can also obtain the cosmological dynamics of branch~\ref{it:coin} from~\ref{it:axi} by setting $z_2=-z_1$, which only affects the curvature term $(z_1+z_2)u^2$ as $z_2$ only appears here. Hence one can obtain the same dynamics as in spatially flat cosmological solutions also in curved spaces by setting $z_1+z_2=0$, regardless of their spatial curvature. We then continue with the branch~\ref{it:vec}, for which the cosmological field equations become significantly more involved, and consist of the metric field equations\\
\begin{subequations}
\begin{align}
3(2z_1 - 2z_3 + z_4 + z_5 + 3z_6 + 4z_7 + 4z_8)u^2 - 3(4z_3 - 2z_4 - 3z_5 - 7z_6 - 18z_7 - 4z_8)\mathcal{H}\mathcal{L} &\nonumber\\
- 9(z_3 + z_7 - 2z_8)\frac{u^4}{\mathcal{L}^2} + 3(z_5 - 3z_6 - 2z_7 + 4z_8)u^2\frac{\mathcal{K}}{\mathcal{L}} - 3(4z_3 - z_5 + 3z_6 + 2z_7 - 4z_8)u^2\frac{\mathcal{H}}{\mathcal{L}} &\nonumber\\
+ 6(z_1 - 2z_6 - 4z_7)\mathcal{H}^2 - 6(z_6 + 2z_7)\mathcal{H}' - 2(3z_4 + 3z_5 + 3z_6 - 6z_7 - 8z_8)\mathcal{H}\mathcal{K} &\nonumber\\
+ 2(3z_6 + 6z_7 + 8z_8)\frac{\mathcal{H}\mathcal{L}'}{\mathcal{L}} - (3z_4 - 6z_6 + 4z_8)\mathcal{K}^2 + 3(z_3 - z_4 - z_5 - 3z_6 - 5z_7 - 2z_8)\mathcal{L}^2 &\nonumber\\
+ 4z_8\frac{2\mathcal{L}\mathcal{L}'' - 3\mathcal{L}'^2}{\mathcal{L}^2} + (6z_6 - 8z_8)\frac{\mathcal{K}\mathcal{L}' - \mathcal{L}\mathcal{K}'}{\mathcal{L}} + 3(2z_4 + z_5 + z_6 - 2z_7 - 4z_8)\mathcal{K}\mathcal{L} &= 2\kappa^2A^2\rho\,,\\
3(z_3 - z_4 - z_5 - 3z_6 - 5z_7 - 2z_8)\mathcal{L}^2 - (4z_3 + z_5 - 3z_6 - 2z_7 + 4z_8)u^2\frac{\mathcal{H}}{\mathcal{L}} &\nonumber\\
- (2z_1 - 2z_3 + z_4 + z_5 + 3z_6 + 4z_7 + 4z_8)u^2 + 2(2z_3 - z_5 - 3z_6 - 10z_7 - 4z_8)\mathcal{L}' &\nonumber\\
- (4z_3 - 6z_4 - 5z_5 - 9z_6 - 14z_7 + 4z_8)\mathcal{H}\mathcal{L} + 3(2z_4 + z_5 + z_6 - 2z_7 - 4z_8)\mathcal{K}\mathcal{L} &\nonumber\\
+ 2(z_5 - 2z_7)\mathcal{K}' - 2(2z_1 + z_4 + z_5)\mathcal{H}' - 2(z_1 + 2z_4 + 2z_5)\mathcal{H}^2 - (z_3 + z_7 - 2z_8)\frac{u^4}{\mathcal{L}^2} &\nonumber\\
- (z_5 - 3z_6 - 2z_7 + 4z_8)u^2\frac{\mathcal{K}}{\mathcal{L}} + 4z_3u^2\frac{\mathcal{L}'}{\mathcal{L}^2} + 2(3z_6 + 2z_7)\frac{\mathcal{H}\mathcal{L}'}{\mathcal{L}} + 2(3z_6 - 4z_8)\frac{\mathcal{K}\mathcal{L}'}{\mathcal{L}} &\nonumber\\
- 2(3z_4 + z_5 - 3z_6 - 2z_7)\mathcal{H}\mathcal{K} - (3z_4 - 6z_6 + 4z_8)\mathcal{K}^2 + 4(z_7 - z_8)\frac{\mathcal{L}'^2}{\mathcal{L}^2} - 4z_7\frac{\mathcal{L}''}{\mathcal{L}} &= 2\kappa^2A^2p\,,
\end{align}
as well as the connection field equations
\begin{align}
4(3z_6 - 4z_8)\mathcal{H}\mathcal{K} + 6(z_3 + z_7 - 2z_8)\frac{u^4}{\mathcal{L}^2} + 3(4z_3 - z_5 + 3z_6 + 2z_7 - 4z_8)u^2\frac{\mathcal{H}}{\mathcal{L}} &\nonumber\\
+ 6(z_6 + 2z_7)(2\mathcal{H}^2 + \mathcal{H}') + 3(4z_3 - 2z_4 - 3z_5 - 7z_6 - 18z_7 - 4z_8)\mathcal{H}\mathcal{L} &\nonumber\\
- 6(z_3 - z_4 - z_5 - 3z_6 - 5z_7 - 2z_8)\mathcal{L}^2 - 3(z_5 - 3z_6 - 2z_7 + 4z_8)u^2\frac{\mathcal{K}}{\mathcal{L}} &\nonumber\\
- 3(2z_4 + z_5 + z_6 - 2z_7 - 4z_8)\mathcal{K}\mathcal{L} + (6z_6 - 8z_8)\mathcal{K}' + 8z_8\frac{\mathcal{L}'^2 - \mathcal{L}\mathcal{L}'' - 2\mathcal{H}\mathcal{L}\mathcal{L}'}{\mathcal{L}^2} &= 2\kappa^2A\left(\omega' + 3\mathcal{H}\omega + 3\mathcal{L}\chi - 3\frac{u^2}{\mathcal{L}}\psi\right)\,,\\
2(z_4 + z_5)(2\mathcal{H}^2 + \mathcal{H}') + 2(z_3 + z_7 - 2z_8)\frac{u^4}{\mathcal{L}^2} - 2(z_3 - z_4 - z_5 - 3z_6 - 5z_7 - 2z_8)\mathcal{L}^2 &\nonumber\\
- (z_5 - 3z_6 - 2z_7 + 4z_8)u^2\frac{\mathcal{L}'}{\mathcal{L}^2} + (4z_3 - 6z_4 - 5z_5 - 9z_6 - 14z_7 + 4z_8)\mathcal{H}\mathcal{L} &\nonumber\\
- (2z_4 + z_5 + z_6 - 2z_7 - 4z_8)(\mathcal{K}\mathcal{L} + \mathcal{L}') + (4z_3 + z_5 - 3z_6 - 2z_7 + 4z_8)u^2\frac{\mathcal{H}}{\mathcal{L}} &\nonumber\\
+ 2(z_4 - z_6)(2\mathcal{H}\mathcal{K} + \mathcal{K}') - (z_5 - 3z_6 - 2z_7 + 4z_8)u^2\frac{\mathcal{K}}{\mathcal{L}} + 2z_6\frac{\mathcal{L}'^2 - \mathcal{L}\mathcal{L}'' - 2\mathcal{H}\mathcal{L}\mathcal{L}'}{\mathcal{L}^2} &= 2\kappa^2A\left(\phi' + 3\mathcal{H}\phi + \mathcal{L}\chi - \frac{u^2}{\mathcal{L}}\psi\right)\,.
\end{align}
\end{subequations}
We see that in contrast to the two previously studied branches, \(\mathcal{L}\) now also appears with second-order time derivatives and the parameter \(z_3\) enters the equations. This holds true also for the remaining flat branches. For the branch~\ref{it:conf} we find the dynamical equations
\begin{subequations}
\begin{align}
2(3z_6 + 6z_7 + 8z_8)\frac{\mathcal{H}\mathcal{L}'}{\mathcal{L}} - 3(4z_3 - 2z_4 - 3z_5 - 7z_6 - 18z_7 - 4z_8)\mathcal{H}\mathcal{L} &\nonumber\\
+ 6(z_1 - 2z_6 - 4z_7)\mathcal{H}^2 - 6(z_6 + 2z_7)\mathcal{H}' - 2(3z_4 + 3z_5 + 3z_6 - 6z_7 - 8z_8)\mathcal{H}\mathcal{K} &\nonumber\\
- (3z_4 - 6z_6 + 4z_8)\mathcal{K}^2 + 3(z_3 - z_4 - z_5 - 3z_6 - 5z_7 - 2z_8)\mathcal{L}^2 &\nonumber\\
+ 4z_8\frac{2\mathcal{L}\mathcal{L}'' - 3\mathcal{L}'^2}{\mathcal{L}^2} + (6z_6 - 8z_8)\frac{\mathcal{K}\mathcal{L}' - \mathcal{L}\mathcal{K}'}{\mathcal{L}} + 3(2z_4 + z_5 + z_6 - 2z_7 - 4z_8)\mathcal{K}\mathcal{L} &= 2\kappa^2A^2\rho\,,\\
3(z_3 - z_4 - z_5 - 3z_6 - 5z_7 - 2z_8)\mathcal{L}^2 - (4z_3 - 6z_4 - 5z_5 - 9z_6 - 14z_7 + 4z_8)\mathcal{H}\mathcal{L} &\nonumber\\
+ 2(2z_3 - z_5 - 3z_6 - 10z_7 - 4z_8)\mathcal{L}' + 3(2z_4 + z_5 + z_6 - 2z_7 - 4z_8)\mathcal{K}\mathcal{L} - 4z_7\frac{\mathcal{L}''}{\mathcal{L}} &\nonumber\\
+ 2(3z_6 + 2z_7)\frac{\mathcal{H}\mathcal{L}'}{\mathcal{L}} - 2(2z_1 + z_4 + z_5)\mathcal{H}' - 2(z_1 + 2z_4 + 2z_5)\mathcal{H}^2 + 2(3z_6 - 4z_8)\frac{\mathcal{K}\mathcal{L}'}{\mathcal{L}} &\nonumber\\
+ 2(z_5 - 2z_7)\mathcal{K}' - 2(3z_4 + z_5 - 3z_6 - 2z_7)\mathcal{H}\mathcal{K} - (3z_4 - 6z_6 + 4z_8)\mathcal{K}^2 + 4(z_7 - z_8)\frac{\mathcal{L}'^2}{\mathcal{L}^2} &= 2\kappa^2A^2p\,,\\
4(3z_6 - 4z_8)\mathcal{H}\mathcal{K} - 6(z_3 - z_4 - z_5 - 3z_6 - 5z_7 - 2z_8)\mathcal{L}^2 &\nonumber\\
+ 6(z_6 + 2z_7)(2\mathcal{H}^2 + \mathcal{H}') + 3(4z_3 - 2z_4 - 3z_5 - 7z_6 - 18z_7 - 4z_8)\mathcal{H}\mathcal{L} &\nonumber\\
- 3(2z_4 + z_5 + z_6 - 2z_7 - 4z_8)\mathcal{K}\mathcal{L} + (6z_6 - 8z_8)\mathcal{K}' + 8z_8\frac{\mathcal{L}'^2 - \mathcal{L}\mathcal{L}'' - 2\mathcal{H}\mathcal{L}\mathcal{L}'}{\mathcal{L}^2} &= 2\kappa^2A\left(\omega' + 3\mathcal{H}\omega + 3\mathcal{L}\chi\right)\,,\\
2(z_4 + z_5)(2\mathcal{H}^2 + \mathcal{H}') - 2(z_3 - z_4 - z_5 - 3z_6 - 5z_7 - 2z_8)\mathcal{L}^2 &\nonumber\\
+ 2(z_4 - z_6)(2\mathcal{H}\mathcal{K} + \mathcal{K}') + (4z_3 - 6z_4 - 5z_5 - 9z_6 - 14z_7 + 4z_8)\mathcal{H}\mathcal{L} &\nonumber\\
- (2z_4 + z_5 + z_6 - 2z_7 - 4z_8)(\mathcal{K}\mathcal{L} + \mathcal{L}') + 2z_6\frac{\mathcal{L}'^2 - \mathcal{L}\mathcal{L}'' - 2\mathcal{H}\mathcal{L}\mathcal{L}'}{\mathcal{L}^2} &= 2\kappa^2A\left(\phi' + 3\mathcal{H}\phi + \mathcal{L}\chi\right)\,,
\end{align}
\end{subequations}
while in the branch~\ref{it:para} they become
\begin{subequations}
\begin{align}
-2(3z_6 + 6z_7 + 8z_8)\frac{\mathcal{H}\mathcal{L}'}{\mathcal{L}} - 3(4z_3 - z_5 + 3z_6 + 2z_7 - 4z_8)\mathcal{H}\mathcal{L} &\nonumber\\
+ 6(z_1 - 2z_6 - 4z_7)\mathcal{H}^2 - 2(3z_4 + 3z_5 + 3z_6 - 6z_7 - 8z_8)\mathcal{H}\mathcal{K} &\nonumber\\
- 6(z_6 + 2z_7)\mathcal{H}' - (3z_4 - 6z_6 + 4z_8)\mathcal{K}^2 - 9(z_3 + z_7 - 2z_8)\mathcal{L}^2 &\nonumber\\
- 4z_8\frac{2\mathcal{L}\mathcal{L}'' - \mathcal{L}'^2}{\mathcal{L}^2} - (6z_6 - 8z_8)\frac{\mathcal{K}\mathcal{L}' + \mathcal{L}\mathcal{K}'}{\mathcal{L}} + 3(z_5 - 3z_6 - 2z_7 + 4z_8)\mathcal{K}\mathcal{L} &= 2\kappa^2A^2\rho\,,\\
2(z_5 - 2z_7)\mathcal{K}' - (z_3 + z_7 - 2z_8)\mathcal{L}^2 - (4z_3 + z_5 - 3z_6 - 2z_7 + 4z_8)\mathcal{H}\mathcal{L} &\nonumber\\
- 4z_3\mathcal{L}' - (z_5 - 3z_6 - 2z_7 + 4z_8)\mathcal{K}\mathcal{L} + 4z_7\frac{\mathcal{L}''}{\mathcal{L}} - (3z_4 - 6z_6 + 4z_8)\mathcal{K}^2 &\nonumber\\
- 2(2z_1 + z_4 + z_5)\mathcal{H}' - 2(z_1 + 2z_4 + 2z_5)\mathcal{H}^2 - 2(3z_6 - 4z_8)\frac{\mathcal{K}\mathcal{L}'}{\mathcal{L}} &\nonumber\\
- 2(3z_6 + 2z_7)\frac{\mathcal{H}\mathcal{L}'}{\mathcal{L}} - 2(3z_4 + z_5 - 3z_6 - 2z_7)\mathcal{H}\mathcal{K} - 4(z_7 + z_8)\frac{\mathcal{L}'^2}{\mathcal{L}^2} &= 2\kappa^2A^2p\,,\\
4(3z_6 - 4z_8)\mathcal{H}\mathcal{K} + 6(z_3 + z_7 - 2z_8)\mathcal{L}^2 + (6z_6 - 8z_8)\mathcal{K}' &\nonumber\\
+ 6(z_6 + 2z_7)(2\mathcal{H}^2 + \mathcal{H}') + 3(4z_3 - z_5 + 3z_6 + 2z_7 - 4z_8)\mathcal{H}\mathcal{L} &\nonumber\\
- 3(z_5 - 3z_6 - 2z_7 + 4z_8)\mathcal{K}\mathcal{L} - 8z_8\frac{\mathcal{L}'^2 - \mathcal{L}\mathcal{L}'' - 2\mathcal{H}\mathcal{L}\mathcal{L}'}{\mathcal{L}^2} &= 2\kappa^2A\left(\omega' + 3\mathcal{H}\omega - 3\mathcal{L}\psi\right)\,,\\
2(z_4 + z_5)(2\mathcal{H}^2 + \mathcal{H}') + 2(z_3 + z_7 - 2z_8)\mathcal{L}^2 - 2z_6\frac{\mathcal{L}'^2 - \mathcal{L}\mathcal{L}'' - 2\mathcal{H}\mathcal{L}\mathcal{L}'}{\mathcal{L}^2} &\nonumber\\
+ 2(z_4 - z_6)(2\mathcal{H}\mathcal{K} + \mathcal{K}') + (4z_3 + z_5 - 3z_6 - 2z_7 + 4z_8)\mathcal{H}\mathcal{L} &\nonumber\\
+ (z_5 - 3z_6 - 2z_7 + 4z_8)(\mathcal{L}' -\mathcal{K}\mathcal{L}) &= 2\kappa^2A\left(\phi' + 3\mathcal{H}\phi - \mathcal{L}\psi\right)\,.
\end{align}
\end{subequations}
Studying the full dynamics arising for the different branches and all possible parameter choices is a major task that would exceed the scope of this article. We therefore restrict ourselves to a few particular classes of theories which are motivated by a field theoretical perspective, which considers general teleparallel quadratic gravity as a theory of three dynamical fields \(h_{\mu\nu}, H_{\mu\nu}, B_{\mu\nu}\) propagating on Minkowski spacetime, where \(h_{\mu\nu}\) is the metric perturbation, while \(H_{\mu\nu}\) and \(B_{\mu\nu}\) are the symmetric and antisymmetric parts of the connection perturbation~\cite{BeltranJimenez:2019odq}. The first class of theories we mention here are those from which the field \(B_{\mu\nu}\), which is a two-form which transforms non-trivially under local Lorentz transformations of the teleparallel connection, is absent. From~\cite{BeltranJimenez:2019odq}, such kind of local Lorentz invariance is realized by the conditions
\begin{equation}
a_2 - 2a_1 = a_3 + 4a_1 = b_3 - b_1 = 0\,.
\end{equation}
Theories satisfying these conditions necessarily have \(z_2 = 0\). Observe that \(z_2\) only enters the field equations~\eqref{eq:cosmoqaxi} in the branch~\ref{it:axi}, where it governs the terms involving the curvature parameter \(u^2\), and is absent in all other branches. This is related to the fact that \(B_{\mu\nu}\) contributes to the axial torsion component, which is non-vanishing only in the branch~\ref{it:axi}, and enforced to vanish by the cosmological symmetry in the other branches. Alternatively, one may consider the weaker condition that \(B_{\mu\nu}\) is a propagating two-form with a \(\mathrm{U}(1)\) gauge symmetry. This is the case if the theory parameters satisfy the conditions
\begin{equation}\label{eq:twoform}
2a_1 + a_2 + a_3 = b_3 - b_1 = 0\,,
\end{equation}
which by themselves have no immediate consequence on the linear combinations \(z_{1,\ldots,8}\) relevant in the cosmological dynamics, but will contribute if they are combined with other conditions as we will see below.

Another type of conditions on the parameters can be obtained that the two fields \(h_{\mu\nu}\) and \(H_{\mu\nu}\) propagate two massless spin-2 fields. This can be obtained if one enhances the diffeomorphism invariance of the theory by another gauge symmetry, such that \(h_{\mu\nu}\) and \(H_{\mu\nu}\) become individually invariant under diffeomorphisms. This is the case for
\begin{equation}\label{eq:difdif}
2c_1 - c_5 = c_1 + c_3 = 2c_1 + c_2 + c_4 = b_1 + b_2 = 0\,,
\end{equation}
and implies \(z_7 = z_8 = 0\). Further demanding that the two spin-2 fields decouple leads to the additional condition \(b_1 = 4c_1\), which then further implies \(z_5 = 0\). Finally, combining these conditions with the condition~\eqref{eq:twoform} of a propagating two-form yields another condition \(z_6 = 0\). Under these conditions, the cosmological dynamics simplify significantly. For the branch~\ref{it:axi} and its flat limit~\ref{it:coin}, the variable \(\mathcal{L}\) decouples completely, and only \(\mathcal{K}\) contributes to the cosmological dynamics. For the remaining three branches, \(\mathcal{L}\) remains in the cosmological field equations, but its equation of motion changes from a second-order to a first-order differential equation, as the second-order derivative terms disappear from the cosmological field equations.

The propagation of two massles spin-2 fields can also be achieved by imposing transverse diffeomorphisms and Weyl symmetry as an additional gauge symmetry. The former gives the condition \(2c_1 + c_2 + c_4 = 0\), while the latter is realized by
\begin{subequations}
\begin{align}
2(2a_1 + a_2)w_H - 2(c_2 + c_4 - 8c_3 - c_5)(w_h - w_H) - b_1(2w_h - 3w_H) - b_2(4w_h - 7w_H) &= 0\,,\\
b_1w_h + 2(b_2 + c_2 + c_4 + 2c_5)(w_h - w_H) - (2a_1 + a_2)w_H &= 0\,,\\
2(c_2 + c_4 - 8c_3 - c_5)(w_h - w_H) - (b_1 + 3b_2)w_H &= 0\,,\\
2(c_2 + c_4 + 2c_5)(w_h - w_H) + b_1w_H &= 0
\end{align}
\end{subequations}
with two constants \(w_h\) and \(w_H\). Imposing these conditions implies that the linear combinations appearing in the cosmological dynamics must satisfy the condition
\begin{equation}
(2z_1 + z_4 + 2z_5 - 2z_7)w_h = (z_5 - 2z_7)w_H\,.
\end{equation}
Imposing in addition the two-form gauge condition~\eqref{eq:twoform} yields a more involved set of equations among the cosmologically relevant parameters, and so we restrict ourselves to two special cases. For \(w_H = 0\) the conditions become
\begin{equation}
2z_1 + z_4 + 2z_5 - 2z_7 = z_5 + z_6 = 3z_7 + 2z_8 = 0\,,
\end{equation}
while in the case \(w_h = w_H \neq 0\) one finds
\begin{equation}
z_1 = z_4 + z_5 = z_6 + 2z_7 = 0\,.
\end{equation}
The former set of conditions does not have any particular relevance for the cosmological dynamics. Under the latter conditions, however, the cosmological dynamics enjoy the curious property that all derivatives of the Hubble parameter \(\mathcal{H}\) disappear from the equation, and it becomes a purely constrained quantity. In this case the dynamics is fully carried by the connection variables \(\mathcal{K}\) and \(\mathcal{L}\). The physical implications of this case, which still has 5 free parameters, as well as those of the other cases discussed above, need to be determined by a more detailed case-by-case analysis, which we defer to future work, and hence conclude our discussion of general teleparallel quadratic gravity at this point.

\subsection{Scalar-teleparallel gravity}\label{ssec:stg}
As the last example, we consider a class of scalar-teleparallel gravity theories, which is defined by the action~\cite{Hohmann:2018ijr,Hohmann:2022mlc}
\begin{equation}\label{eq:stgaction}
S_{\text{g}} = \frac{1}{2\kappa^2}\dd^4x\sqrt{-g}\int\left[-\mathcal{A}(\Phi)G - \mathcal{B}(\Phi)g^{\mu\nu}\partial_{\mu}\Phi\partial_{\nu}\Phi + \mathcal{C}(\Phi)(2T_{\nu}{}^{\nu\mu} - Q^{\mu\nu}{}_{\nu} + Q_{\nu}{}^{\nu\mu})\partial_{\mu}\Phi - 2\kappa^2\mathcal{V}(\Phi)\right]\,,
\end{equation}
where \(\Phi\) denotes a scalar field, and \(\mathcal{A}, \mathcal{B}, \mathcal{C}, \mathcal{V}\) are free functions, whose choice determines a particular theory within this class. The field equations are given by the metric equation
\begin{multline}\label{eq:stgmetfield}
\mathcal{A}\lc{R}_{\mu\nu} - \frac{\mathcal{A}}{2}\lc{R}g_{\mu\nu} + \mathcal{C}\lc{\nabla}_{\mu}\lc{\nabla}_{\nu}\Phi + (\mathcal{A}' + \mathcal{C})\left(\lc{\nabla}_{(\mu}\Phi M^{\rho}{}_{\nu)\rho} - M^{\rho}{}_{(\mu\nu)}\lc{\nabla}_{\rho}\Phi + M^{[\rho\sigma]}{}_{\sigma}\lc{\nabla}_{\rho}\Phi g_{\mu\nu}\right)\\
- (\mathcal{B} - \mathcal{C}')\lc{\nabla}_{\mu}\Phi\lc{\nabla}_{\nu}\Phi + \left[\left(\frac{\mathcal{B}}{2} - \mathcal{C}'\right)\lc{\nabla}_{\rho}\Phi\lc{\nabla}^{\rho}\Phi - \mathcal{C}\lc{\nabla}_{\rho}\lc{\nabla}^{\rho}\Phi + \kappa^2\mathcal{V}\right]g_{\mu\nu} = \kappa^2\Theta_{\mu\nu}\,,
\end{multline}
the connection equation
\begin{equation}\label{eq:stgconnfield}
(\mathcal{A}' + \mathcal{C})\left[M^{\nu\rho}{}_{\rho}\lc{\nabla}_{\mu}\Phi + M^{\rho}{}_{\mu\rho}\lc{\nabla}^{\nu}\Phi - (M^{\nu\rho}{}_{\mu} + M^{\rho}{}_{\mu}{}^{\nu})\lc{\nabla}_{\rho}\Phi\right] = 2\kappa^2(\nabla_{\tau}H_{\mu}{}^{\nu\tau} - M^{\omega}{}_{\tau\omega}H_{\mu}{}^{\nu\tau})
\end{equation}
and the scalar field equation
\begin{equation}
-2\mathcal{B}\lc{\nabla}_{\mu}\lc{\nabla}^{\mu}\Phi - \mathcal{B}'\lc{\nabla}_{\mu}\Phi\lc{\nabla}^{\mu}\Phi + \mathcal{C}B + \mathcal{A}'G + 2\kappa^2\mathcal{V}' = 0\,.
\end{equation}
Note that in the case \(\mathcal{C} = -\mathcal{A}'\) this class of theories reduces to the well-studied class of scalar-curvature theories~\cite{Hohmann:2018ijr,Hohmann:2022mlc}.

We then take a look at the cosmological dynamics. For the branch~\ref{it:axi} the metric and scalar field equations become
\begin{subequations}
\begin{align}
6\mathcal{A}(\mathcal{H}^2 + u^2) - \mathcal{B}\Phi'^2 - 6\mathcal{C}\mathcal{H}\Phi' - 2\kappa^2A^2\mathcal{V} &= 2\kappa^2A^2\rho\,,\\
-2\mathcal{A}(2\mathcal{H}' + \mathcal{H}^2 + u^2) + (2\mathcal{C}' - \mathcal{B})\Phi'^2 - 2(2\mathcal{A}' + \mathcal{C})\mathcal{H}\Phi' + 2\mathcal{C}\Phi'' + 2\kappa^2A^2\mathcal{V} &= 2\kappa^2A^2p\,,\\
6\mathcal{C}\mathcal{H}' + 6(2\mathcal{C} + \mathcal{A}')\mathcal{H}^2 - 6\mathcal{A}'u^2 + 2\mathcal{B}(\Phi'' + 2\mathcal{H}\Phi') + \mathcal{B}'\Phi'^2 + 2\kappa^2A^2\mathcal{V}' &= 0\,.
\end{align}
\end{subequations}
The left hand side of the connection field equations vanishes identically, and so these equations are accompanied by the hypermomentum conservation equations
\begin{equation}
\omega' + 3\mathcal{H}\omega = \phi' + 3\mathcal{H}\phi = 0\,.
\end{equation}
By taking the limit \(u \to 0\), this branch reduces to the branch~\ref{it:coin}, and the field equations become
\begin{subequations}
\begin{align}
6\mathcal{A}\mathcal{H}^2 - \mathcal{B}\Phi'^2 - 6\mathcal{C}\mathcal{H}\Phi' - 2\kappa^2A^2\mathcal{V} &= 2\kappa^2A^2\rho\,,\\
-2\mathcal{A}(2\mathcal{H}' + \mathcal{H}^2) + (2\mathcal{C}' - \mathcal{B})\Phi'^2 - 2(2\mathcal{A}' + \mathcal{C})\mathcal{H}\Phi' + 2\mathcal{C}\Phi'' + 2\kappa^2A^2\mathcal{V} &= 2\kappa^2A^2p\,,\\
6\mathcal{C}\mathcal{H}' + 6(2\mathcal{C} + \mathcal{A}')\mathcal{H}^2 + 2\mathcal{B}(\Phi'' + 2\mathcal{H}\Phi') + \mathcal{B}'\Phi'^2 + 2\kappa^2A^2\mathcal{V}' &= 0\,.
\end{align}
\end{subequations}
Note that in these two cases the two functions \(\mathcal{K}\) and \(\mathcal{L}\) describing the cosmologically symmetric connection do not enter the cosmological dynamics. One finds that the obtained cosmological dynamics agree with those of a class of scalar-torsion theories of gravity for the axial branch of metric teleparallel cosmology and its flat limit~\cite{Hohmann:2020zre,Hohmann:2018ijr,Hohmann:2020zre}.

We continue with the spatially curved branch~\ref{it:vec}. In this case the metric field equations become
\begin{subequations}\label{eq:cosmostgvecm}
\begin{align}
6\mathcal{A}(\mathcal{H}^2 + u^2) - \mathcal{B}\Phi'^2 + 3(\mathcal{A}' + \mathcal{C})\left(\mathcal{L} + \frac{u^2}{\mathcal{L}}\right)\Phi' - 6\mathcal{C}\mathcal{H}\Phi' - 2\kappa^2A^2\mathcal{V} &= 2\kappa^2A^2\rho\,,\\
-2\mathcal{A}(2\mathcal{H}' + \mathcal{H}^2 + u^2) + (2\mathcal{C}' - \mathcal{B})\Phi'^2 - 2(2\mathcal{A}' + \mathcal{C})\mathcal{H}\Phi' + (\mathcal{A}' + \mathcal{C})\left(3\mathcal{L} - \frac{u^2}{\mathcal{L}}\right)\Phi' + 2\mathcal{C}\Phi'' + 2\kappa^2A^2\mathcal{V} &= 2\kappa^2A^2p\,.
\end{align}
\end{subequations}
The connection equations are also non-trivial in this case and read
\begin{subequations}\label{eq:cosmostgvecc}
\begin{align}
3(\mathcal{A}' + \mathcal{C})\left(\mathcal{L} + \frac{u^2}{\mathcal{L}}\right)\Phi' &= 2\kappa^2A\left(\omega' + 3\mathcal{H}\omega + 3\mathcal{L}\chi - 3\frac{u^2}{\mathcal{L}}\psi\right)\,,\\
(\mathcal{A}' + \mathcal{C})\left(\mathcal{L} + \frac{u^2}{\mathcal{L}}\right)\Phi' &= 2\kappa^2A\left(\phi' + 3\mathcal{H}\phi + \mathcal{L}\chi - \frac{u^2}{\mathcal{L}}\psi\right)\,.
\end{align}
\end{subequations}
Finally, the scalar equation takes the form
\begin{equation}\label{eq:cosmostgvecs}
6\mathcal{C}\mathcal{H}' + 6(2\mathcal{C} + \mathcal{A}')\mathcal{H}^2 - 6\mathcal{A}'u^2 - 3(\mathcal{A}' + \mathcal{C})\left[2\mathcal{H}\left(\mathcal{L} - \frac{u^2}{\mathcal{L}}\right) + \frac{\mathcal{L}'}{\mathcal{L}}\left(\mathcal{L} + \frac{u^2}{\mathcal{L}}\right)\right] + 2\mathcal{B}(\Phi'' + 2\mathcal{H}\Phi') + \mathcal{B}'\Phi'^2 + 2\kappa^2A^2\mathcal{V}' = 0\,.
\end{equation}
Before analyzing these equations, we also list the remaining spatially flat branches. In the branch~\ref{it:conf}, the field equations become
\begin{subequations}
\begin{align}
6\mathcal{A}(\mathcal{H}^2 + u^2) - \mathcal{B}\Phi'^2 + 3(\mathcal{A}' + \mathcal{C})\mathcal{L}\Phi' - 6\mathcal{C}\mathcal{H}\Phi' - 2\kappa^2A^2\mathcal{V} &= 2\kappa^2\rho\,,\\
-2\mathcal{A}(2\mathcal{H}' + \mathcal{H}^2 + u^2) + (2\mathcal{C}' - \mathcal{B})\Phi'^2 - 2(2\mathcal{A}' + \mathcal{C})\mathcal{H}\Phi' + 3(\mathcal{A}' + \mathcal{C})\mathcal{L}\Phi' + 2\mathcal{C}\Phi'' + 2\kappa^2A^2\mathcal{V} &= 2\kappa^2p\\
3(\mathcal{A}' + \mathcal{C})\mathcal{L}\Phi' &= 2\kappa^2A(\omega' + 3\mathcal{H}\omega + 3\mathcal{L}\chi)\,,\\
(\mathcal{A}' + \mathcal{C})\mathcal{L}\Phi' &= 2\kappa^2A(\phi' + 3\mathcal{H}\phi + \mathcal{L}\chi)\,,\\
6\mathcal{C}\mathcal{H}' + 6(2\mathcal{C} + \mathcal{A}')\mathcal{H}^2 - 6\mathcal{A}'u^2 - 3(\mathcal{A}' + \mathcal{C})(\mathcal{L}' + 2\mathcal{H}\mathcal{L}) + 2\mathcal{B}(\Phi'' + 2\mathcal{H}\Phi') + \mathcal{B}'\Phi'^2 + 2\kappa^2A^2\mathcal{V}' &= 0\,,
\end{align}
\end{subequations}
while for the branch~\ref{it:para} we find
\begin{subequations}
\begin{align}
6\mathcal{A}(\mathcal{H}^2 + u^2) - \mathcal{B}\Phi'^2 + 3(\mathcal{A}' + \mathcal{C})\mathcal{L}\Phi' - 6\mathcal{C}\mathcal{H}\Phi' - 2\kappa^2A^2\mathcal{V} &= 2\kappa^2\rho\,,\\
-2\mathcal{A}(2\mathcal{H}' + \mathcal{H}^2 + u^2) + (2\mathcal{C}' - \mathcal{B})\Phi'^2 - 2(2\mathcal{A}' + \mathcal{C})\mathcal{H}\Phi' - (\mathcal{A}' + \mathcal{C})\mathcal{L}\Phi' + 2\mathcal{C}\Phi'' + 2\kappa^2A^2\mathcal{V} &= 2\kappa^2p\\
3(\mathcal{A}' + \mathcal{C})\mathcal{L}\Phi' &= 2\kappa^2A(\omega' + 3\mathcal{H}\omega - 3\mathcal{L}\psi)\,,\\
(\mathcal{A}' + \mathcal{C})\mathcal{L}\Phi' &= 2\kappa^2A(\phi' + 3\mathcal{H}\phi - \mathcal{L}\psi)\,,\\
6\mathcal{C}\mathcal{H}' + 6(2\mathcal{C} + \mathcal{A}')\mathcal{H}^2 - 6\mathcal{A}'u^2 + 3(\mathcal{A}' + \mathcal{C})(\mathcal{L}' + 2\mathcal{H}\mathcal{L}) + 2\mathcal{B}(\Phi'' + 2\mathcal{H}\Phi') + \mathcal{B}'\Phi'^2 + 2\kappa^2A^2\mathcal{V}' &= 0\,.
\end{align}
\end{subequations}
Note that in all three cases only the parameter function \(\mathcal{L}\) contributes to the field equations, while \(\mathcal{K}\) is absent.
%For constant scalar field, $\Phi'=0$, the action reduces to the GTEGR action and the equations of motion become the Friedmann equations up to a constant rescaling of the gravitational constant, mediated by \(\mathcal{A}\), which can be absorbed into \(\kappa^2\), as well as a cosmological constant determined by the potential \(\mathcal{V}\). Setting $\Phi=$const in the full equations of motion however yields to an inconsistent scalar field equation, since it reduces to $\mathcal{C}B+\mathcal{A}G'+2\kappa^2\mathcal{V}'=0$, which in general cannot be solved for the constant $\Phi$, unless also \(\mathcal{C}B + \mathcal{A}G'\) is constant.

To study the dynamics of the three branches~\ref{it:vec}, \ref{it:para} and~\ref{it:conf}, we restrict ourselves to the case of vanishing hypermomentum, and start with the connection equations. For the spatially curved branch~\ref{it:vec}, these take the form~\eqref{eq:cosmostgvecc} and can be solved by any of the following solutions:
\begin{enumerate}
\item
For theories which satisfy the condition \(\mathcal{A}' + \mathcal{C} = 0\) one can combine the terms containing $\mathcal{A}$ and $\mathcal{C}$ in the action to $\mathcal{A}(-G+B)=\mathcal{A}\accentset{\circ}{R}$ by integrating by parts, and the theory reduces to an equivalent of scalar-curvature gravity~\cite{Hohmann:2018ijr}. The teleparallel connection variable \(\mathcal{L}\) does not appear in the field equations, and the connection equations~\eqref{eq:cosmostgvecc} of motion become trivial, i.e., the left hand side of these equations vanishes, and so they are solved identically for vanishing hypermomentum. The same holds also for the flat branches~\ref{it:conf} and~\ref{it:para}.
\item
The field equations~\eqref{eq:cosmostgvecc} are also solved if the scalar field is constant, \(\Phi' = 0\). Also in this case \(\mathcal{L}\) disappears from the metric field equations~\eqref{eq:cosmostgvecm}, and they reduce to the Friedmann equations, up to a constant rescaling of the gravitational constant, mediated by \(\mathcal{A}\), which can be absorbed into \(\kappa^2\), as well as a cosmological constant determined by the potential \(\mathcal{V}\). The remaining scalar field equation~\eqref{eq:cosmostgvecs}, which is a first-order differential equation in \(\mathcal{L}\), can then be solved for \(\mathcal{L}\). Also these findings apply in full analogy to the flat branches~\ref{it:conf} and~\ref{it:para}.
\item
For \(u^2 < 0\), corresponding to negative spatial curvature, the connection field equations~\eqref{eq:cosmostgvecc} can also be solved by \(\mathcal{L} = \pm iu\). Choosing the lower sign, both the metric field equations~\eqref{eq:cosmostgvecm} and the scalar field equation~\eqref{eq:cosmostgvecs} reduce to the dynamics of the vector branch in scalar-torsion gravity~\cite{Hohmann:2020zre,Hohmann:2018ijr,Hohmann:2020zre}, while for the upper sign one has a trivial substitution \(u \mapsto -u\). The same statement holds also for the spatially flat branches~\ref{it:conf} and~\ref{it:para}, where in this case one sets \(\mathcal{L} = 0\) and the cosmological dynamics reduce to that of scalar-torsion theory in the limit \(u = 0\) of vanishing spatial curvature.
\end{enumerate}

In summary, we find that for all branches of cosmologically symmetric teleparallel geometries the cosmological dynamics of the class of scalar-teleparallel gravity theories defined by the action~\eqref{eq:stgaction} reduce to those of scalar-torsion gravity~\cite{Hohmann:2018ijr}, or one of its special cases, which are equivalent to scalar-curvature gravity or general relativity with a cosmological constant. In order to obtain a non-trivial contribution also from the nonmetricity to the cosmological dynamics, one needs to study more classes of general scalar-teleparallel theories. We leave such studies for future work.

\section{Example solution}\label{sec:solution}
We have already seen some example solutions of the connection equations of motion in $f(G)$ gravity in the absence of hypermomentum. Here we want to show another simple case exemplifying effects of the connection and hypermomentum on the metric. We consider the general quadratic gravity in the simple branch~\ref{it:axi}. One can show that one can find the exact solution of the connection equations
\begin{align}
    \mathcal{K}&=\frac{\mathcal{K}_0}{A^2}+\frac{(-3z_6(z_6+2z_7)+4z_8(z_4+z_5))\mathcal{H}+A(4z_8\phi-z_6\omega)}{3z_6^2-4z_4z_8}\,,\\
    \mathcal{L}&=\frac{\mathcal{L}_0}{A^2}+\frac{(-3z_5z_6+6z_4z_7)\mathcal{H}+A(-3z_6\phi+z_4\omega)}{3z_6^2-4z_4z_8}\,,
\end{align}
with $\mathcal{K}_0$ and $\mathcal{L}_0$ integration constants, and we have to assume here $3z_6^2-4z_4z_8\neq0$. For $3z_6^2-4z_4z_8=0$ only the combination $z_4\mathcal{K}+z_6\mathcal{L}$ appears in the connection equations of motion, so only this sum can be determined from them. We also absorb here $\kappa^2$ in the hypermomentum variables $\phi$ and $\omega$. Plugging the connection functions in the metric equations of motion yields
\begin{align}
    2\kappa^2\rho&=6\alpha\frac{\mathcal{H}^2}{A^2}+6(z_1+z_2)\frac{u^2}{A^2}+\rho_\Gamma+\rho_H+\frac{2}{A}(\omega'+3\mathcal{H}\omega)\,,\\
    2\kappa^2 p&=2\alpha(\mathcal{H}'+\mathcal{H}^2)-2(z_1+z_2)\frac{u^2}{A^2}+\rho_\Gamma+\rho_H-\frac{2}{A}(\tilde\phi'+3\mathcal{H}\tilde\phi)\,,
\end{align}
with
\begin{align}
    \alpha&=z_1-\frac{4z_4^2z_8+2z_5(-3z_6^2-6z_6z_7+2z_5z_8)+z_4(-3z_6^2+12z_7^2+8z_5z_8)}{6z_6^2-8z_4z_8}\,,\\
    \rho_\Gamma&=-\frac{3z_4\mathcal{K}_0^2+6z_6\mathcal{K}_0\mathcal{L}_0+4z_8\mathcal{L}_0^2}{A^6}\,,\\
    \rho_H&=\frac{6\mathcal{K}_0\phi+2\mathcal{L}_0\omega}{A^3}+\frac{12z_8\phi^2-6z_6\phi\omega+z_4\omega^2}{3z_6^2-4z_4z_8}\,,\\
    \tilde\phi&=\frac{(6z_6z_7-4z_5z_8)\phi+(z_5z_6-2z_4z_7)\omega}{3z_6^2-4z_4z_8}\,.
\end{align}
The constant $\alpha$ determines the coupling to matter, which should be unity to obtain the same coupling as in GR. Spatial curvature again enters as an effective $(z_1+z_2)u^2$. We also find a contribution coming from the connection, $\rho_\Gamma\propto A^{-6}$, which shows that the connection behaves like stiff matter $p_\Gamma=\rho_\Gamma$. Note that the sign of the energy density depends on the signs of the $z_i$, and one can achieve $\rho_\Gamma<0$. Such stiff matter with negative energy density may be used to facilitate a bouncing universe. Hypermomentum enters in a more complicated way.
%, however we still need to assure the continuity equation $\rho'+3\mathcal{H}(\rho+p)=0$ of matter. A simple solution is if $\omega$ and $\phi$ fulfill the separate conservation equations $\omega'+3\mathcal{H}\omega=\phi'+3\mathcal{H}\phi=0$, leading to $\omega\propto A^{-3}$ and $\phi\propto A^{-3}$. Then $\rho_H\propto A^{-6}$, so hypermomentum also behaves like stiff matter.
%
%Another interesting solution is obtained for the case $z_1=1$ and $z_4=z_5=z_7=z_8=0$ but $z_6\neq 0$. Then $\alpha=1$ and $\tilde\phi=0$, and we can also set $\phi=\mathcal{L}_0=0$ as well leading to $\rho_H=0$. Defining $\Omega=\omega'+3\mathcal{H}\omega$ one finds that the continuity equation implies \MH{\textbf{Should be $\Omega = 0$?}}
%\begin{equation}
%    \Omega'+2\mathcal{H}\Omega=0\ ,
%\end{equation}
%with solution $\Omega=\Omega_0 A^{-2}$ with $H_0$ an integration constant. From this we find
%\begin{equation}
%    \omega=\frac{\omega_0}{A^3}+\frac{\Omega_0}{A^3}\int\dd\eta\,A\ ,
%\end{equation}
%with $\eta$ conformal time. But now we find that the only contribution to the energy density is
%\begin{equation}
%    \frac{\Omega}{A}=\frac{\Omega_0}{A^3}\ ,
%\end{equation}
%and no pressure contribution. Hence the hypermomentum in the form of $\Omega$ now enters as dark matter.

\section{Conclusion}\label{sec:conclusion}
We have derived the most general class of homogeneous and isotropic teleparallel geometries, defined by a metric and a flat, affine connection, which are invariant under spatial rotations and translations. We find that there are five branches of such geometries, two of which exhibit a non-vanishing spatial curvature for the metric Levi-Civita connection, while the remaining three spatially flat cases arise as particular limits from the spatially curved cases. We have also shown how these branches are related to the more restricted classes or metric and symmetric teleparallel geometries, in which nonmetricity or torsion are imposed to vanish. Our findings show that in addition to the lapse and scale factor appearing in the homogeneous and isotropic Robertson-Walker metric, the flat affine connection is described by two further functions of time, which may participate in the cosmological dynamics for a suitable gravity theory which couples to these degrees of freedom.

We have then applied our findings to a number of general teleparallel gravity theories and derived their cosmological dynamics. In particular, we have considered the $f(G)$ class of theories, general teleparallel quadratic gravity and a simple class of scalar-teleparallel gravity theories. We have seen that for the different branches of cosmologically symmetric teleparallel geometries, which are distinct only by the flat, affine connection, one finds, in general, qualitatively different cosmological dynamics, such as a different number of dynamical functions appearing in the cosmological field equations or a different differential order of these equations. Further, our findings show that for the $f(G)$ and scalar-teleparallel theories the cosmological dynamics fully reduces to that of related metric teleparallel or curvature based gravity theories, in which nonmetricity and possibly torsion are absent, such that the results found for the cosmological dynamics of these simpler theories also apply to their general teleparallel counterparts as follows:
\begin{enumerate}
\item
Already without imposing cosmological symmetry, $f(G)$ gravity reduces to general relativity with a cosmological constant (GR$\Lambda$) for \(f'' \equiv 0\), while the scalar-teleparallel theories reduce to scalar-curvature gravity (SCG) for \(\mathcal{A}' + \mathcal{C} \equiv 0\).
\item
In the branch~\ref{it:axi}, the dynamics reduces to that of the metric teleparallel counterpart ($f(T)$ gravity or scalar-torsion gravity (STG)) in the ``axial'' branch~\cite{Hohmann:2020zre}, and to its flat limit in the branch~\ref{it:coin}. The latter also holds also for vanishing hypermomentum in the branches~\ref{it:para} and~\ref{it:conf} for \(\mathcal{L} = 0\), while one finds the ``vector'' branch of metric teleparallel solutions for the branch~\ref{it:vec} and \(\mathcal{L} = \pm iu\).
\item
For the spatially curved branch~\ref{it:vec} and the two flat branches~\ref{it:para} and~\ref{it:conf}, one finds that if \(\mathcal{L}\) does not take the values given in the preceding item, the connection equations with vanishing hypermomentum inevitably yield a solution for which the metric field equations reduce to that of general relativity with a cosmological constant.
\end{enumerate}
It follows that the only possibility to obtain new dynamics in the aforementioned classes of theories is to consider a non-trivial coupling between matter and the teleparallel connection, leading to non-vanishing hypermomentum. For more general theories, new dynamics is obtained. As an explicit example, we have shown this for the class of general teleparallel quadratic gravity theories, but one may also consider theories with a more general dependence of the Lagrangian on the teleparallel geometry or more general scalar field couplings.

Our results allow for several directions of further research. Starting from the homogeneous and isotropic teleparallel geometries we have determined, one can derive and study the cosmological dynamics of further general teleparallel gravity theories. Also among those classes of theories whose cosmological field equations we have derived in this article we find several classes of general teleparallel quadratic gravity theories whose cosmological dynamics qualitatively differs from the previously studied metric and symmetric teleparallel theories, and deserves further attention. Finally, going beyond the cosmological background dynamics one may consider perturbations of the teleparallel geometry around this background and study their dynamics. For a general flat connection one finds that the perturbations are of the form
\begin{equation}
    \delta\Gamma^\alpha{}_{\mu\nu}=\nabla_\nu\delta\Lambda^\alpha{}_\mu\ ,
\end{equation}
with $\delta\Lambda^\alpha{}_\mu$ an arbitrary matrix with 16 entries, in addition to the ten perturbation variables of the metric. Even after gauging away four of these variables one is left with 22 perturbation variables, leading to a much more cumbersome analysis compared to the mere three free functions of the cosmological background metric and connection. An important question to be addressed in these studies is whether the so-called strong coupling problem for linear perturbations around highly symmetric background, which has been found in metric teleparallel gravity theories~\cite{Golovnev:2020zpv,Golovnev:2018wbh,Golovnev:2020nln,Jimenez:2020ofm,Blagojevic:2020dyq,Guzman:2019oth,Bahamonde:2022ohm}, is also present in general teleparallel gravity. FLRW solutions based on non-metricity seem to provide a promising route for this purpose \cite{BeltranJimenez:2019tme}. We leave this question for future investigations.

\begin{acknowledgments}
LH is supported by funding from the European Research Council (ERC) under the European Unions Horizon 2020 research and innovation programme grant agreement No 801781 and by the Swiss National Science Foundation grant 179740.
MH gratefully acknowledges the full financial support by the Estonian Research Council through the Personal Research Funding project PRG356 and by the European Regional Development Fund through the Center of Excellence TK133 ``The Dark Side of the Universe''. The authors acknowledge networking support by the COST Actions CA18108 and CA21136.
\end{acknowledgments}

\bibliography{telecosmo}

\begin{thebibliography}{27}%
\makeatletter
\providecommand \@ifxundefined [1]{%
 \@ifx{#1\undefined}
}%
\providecommand \@ifnum [1]{%
 \ifnum #1\expandafter \@firstoftwo
 \else \expandafter \@secondoftwo
 \fi
}%
\providecommand \@ifx [1]{%
 \ifx #1\expandafter \@firstoftwo
 \else \expandafter \@secondoftwo
 \fi
}%
\providecommand \natexlab [1]{#1}%
\providecommand \enquote  [1]{``#1''}%
\providecommand \bibnamefont  [1]{#1}%
\providecommand \bibfnamefont [1]{#1}%
\providecommand \citenamefont [1]{#1}%
\providecommand \href@noop [0]{\@secondoftwo}%
\providecommand \href [0]{\begingroup \@sanitize@url \@href}%
\providecommand \@href[1]{\@@startlink{#1}\@@href}%
\providecommand \@@href[1]{\endgroup#1\@@endlink}%
\providecommand \@sanitize@url [0]{\catcode `\\12\catcode `\$12\catcode
  `\&12\catcode `\#12\catcode `\^12\catcode `\_12\catcode `\%12\relax}%
\providecommand \@@startlink[1]{}%
\providecommand \@@endlink[0]{}%
\providecommand \url  [0]{\begingroup\@sanitize@url \@url }%
\providecommand \@url [1]{\endgroup\@href {#1}{\urlprefix }}%
\providecommand \urlprefix  [0]{URL }%
\providecommand \Eprint [0]{\href }%
\providecommand \doibase [0]{https://doi.org/}%
\providecommand \selectlanguage [0]{\@gobble}%
\providecommand \bibinfo  [0]{\@secondoftwo}%
\providecommand \bibfield  [0]{\@secondoftwo}%
\providecommand \translation [1]{[#1]}%
\providecommand \BibitemOpen [0]{}%
\providecommand \bibitemStop [0]{}%
\providecommand \bibitemNoStop [0]{.\EOS\space}%
\providecommand \EOS [0]{\spacefactor3000\relax}%
\providecommand \BibitemShut  [1]{\csname bibitem#1\endcsname}%
\let\auto@bib@innerbib\@empty
%</preamble>
\bibitem [{\citenamefont {Aghanim}\ \emph {et~al.}(2020)\citenamefont {Aghanim}
  \emph {et~al.}}]{Planck:2018vyg}%
  \BibitemOpen
  \bibfield  {author} {\bibinfo {author} {\bibfnamefont {N.}~\bibnamefont
  {Aghanim}} \emph {et~al.} (\bibinfo {collaboration} {Planck}),\ }\bibfield
  {title} {\bibinfo {title} {{Planck 2018 results. VI. Cosmological
  parameters}},\ }\href {https://doi.org/10.1051/0004-6361/201833910}
  {\bibfield  {journal} {\bibinfo  {journal} {Astron. Astrophys.}\ }\textbf
  {\bibinfo {volume} {641}},\ \bibinfo {pages} {A6} (\bibinfo {year} {2020})},\
  \bibinfo {note} {[Erratum: Astron.Astrophys. 652, C4 (2021)]},\ \Eprint
  {https://arxiv.org/abs/1807.06209} {arXiv:1807.06209 [astro-ph.CO]}
  \BibitemShut {NoStop}%
\bibitem [{\citenamefont {Di~Valentino}\ \emph {et~al.}(2021)\citenamefont
  {Di~Valentino}, \citenamefont {Mena}, \citenamefont {Pan}, \citenamefont
  {Visinelli}, \citenamefont {Yang}, \citenamefont {Melchiorri}, \citenamefont
  {Mota}, \citenamefont {Riess},\ and\ \citenamefont
  {Silk}}]{DiValentino:2021izs}%
  \BibitemOpen
  \bibfield  {author} {\bibinfo {author} {\bibfnamefont {E.}~\bibnamefont
  {Di~Valentino}}, \bibinfo {author} {\bibfnamefont {O.}~\bibnamefont {Mena}},
  \bibinfo {author} {\bibfnamefont {S.}~\bibnamefont {Pan}}, \bibinfo {author}
  {\bibfnamefont {L.}~\bibnamefont {Visinelli}}, \bibinfo {author}
  {\bibfnamefont {W.}~\bibnamefont {Yang}}, \bibinfo {author} {\bibfnamefont
  {A.}~\bibnamefont {Melchiorri}}, \bibinfo {author} {\bibfnamefont {D.~F.}\
  \bibnamefont {Mota}}, \bibinfo {author} {\bibfnamefont {A.~G.}\ \bibnamefont
  {Riess}},\ and\ \bibinfo {author} {\bibfnamefont {J.}~\bibnamefont {Silk}},\
  }\bibfield  {title} {\bibinfo {title} {{In the realm of the Hubble
  tension\textemdash{}a review of solutions}},\ }\href
  {https://doi.org/10.1088/1361-6382/ac086d} {\bibfield  {journal} {\bibinfo
  {journal} {Class. Quant. Grav.}\ }\textbf {\bibinfo {volume} {38}},\ \bibinfo
  {pages} {153001} (\bibinfo {year} {2021})},\ \Eprint
  {https://arxiv.org/abs/2103.01183} {arXiv:2103.01183 [astro-ph.CO]}
  \BibitemShut {NoStop}%
\bibitem [{\citenamefont {Saridakis}\ \emph {et~al.}(2021)\citenamefont
  {Saridakis} \emph {et~al.}}]{Saridakis:2021lqd}%
  \BibitemOpen
  \bibfield  {author} {\bibinfo {author} {\bibfnamefont {E.~N.}\ \bibnamefont
  {Saridakis}} \emph {et~al.} (\bibinfo {collaboration} {CANTATA}),\ }\bibfield
   {title} {\bibinfo {title} {{Modified Gravity and Cosmology: An Update by the
  CANTATA Network}},\ }\href@noop {} {\  (\bibinfo {year} {2021})},\ \Eprint
  {https://arxiv.org/abs/2105.12582} {arXiv:2105.12582 [gr-qc]} \BibitemShut
  {NoStop}%
%%CITATION = ARXIV:2105.12582;%%
\bibitem [{\citenamefont {Heisenberg}(2019)}]{Heisenberg:2018vsk}%
  \BibitemOpen
  \bibfield  {author} {\bibinfo {author} {\bibfnamefont {L.}~\bibnamefont
  {Heisenberg}},\ }\bibfield  {title} {\bibinfo {title} {{A systematic approach
  to generalisations of General Relativity and their cosmological
  implications}},\ }\href {https://doi.org/10.1016/j.physrep.2018.11.006}
  {\bibfield  {journal} {\bibinfo  {journal} {Phys. Rept.}\ }\textbf {\bibinfo
  {volume} {796}},\ \bibinfo {pages} {1} (\bibinfo {year} {2019})},\ \Eprint
  {https://arxiv.org/abs/1807.01725} {arXiv:1807.01725 [gr-qc]} \BibitemShut
  {NoStop}%
\bibitem [{\citenamefont {Jim{\'e}nez}\ \emph {et~al.}(2019)\citenamefont
  {Jim{\'e}nez}, \citenamefont {Heisenberg},\ and\ \citenamefont
  {Koivisto}}]{BeltranJimenez:2019tjy}%
  \BibitemOpen
  \bibfield  {author} {\bibinfo {author} {\bibfnamefont {J.~B.}\ \bibnamefont
  {Jim{\'e}nez}}, \bibinfo {author} {\bibfnamefont {L.}~\bibnamefont
  {Heisenberg}},\ and\ \bibinfo {author} {\bibfnamefont {T.~S.}\ \bibnamefont
  {Koivisto}},\ }\bibfield  {title} {\bibinfo {title} {{The Geometrical Trinity
  of Gravity}},\ }\href {https://doi.org/10.3390/universe5070173} {\bibfield
  {journal} {\bibinfo  {journal} {Universe}\ }\textbf {\bibinfo {volume} {5}},\
  \bibinfo {pages} {173} (\bibinfo {year} {2019})},\ \Eprint
  {https://arxiv.org/abs/1903.06830} {arXiv:1903.06830 [hep-th]} \BibitemShut
  {NoStop}%
%%CITATION = ARXIV:1903.06830;%%
\bibitem [{\citenamefont {Beltr\'an~Jim\'enez}\ \emph
  {et~al.}(2020{\natexlab{a}})\citenamefont {Beltr\'an~Jim\'enez},
  \citenamefont {Heisenberg}, \citenamefont {Iosifidis}, \citenamefont
  {Jim\'enez-Cano},\ and\ \citenamefont {Koivisto}}]{BeltranJimenez:2019odq}%
  \BibitemOpen
  \bibfield  {author} {\bibinfo {author} {\bibfnamefont {J.}~\bibnamefont
  {Beltr\'an~Jim\'enez}}, \bibinfo {author} {\bibfnamefont {L.}~\bibnamefont
  {Heisenberg}}, \bibinfo {author} {\bibfnamefont {D.}~\bibnamefont
  {Iosifidis}}, \bibinfo {author} {\bibfnamefont {A.}~\bibnamefont
  {Jim\'enez-Cano}},\ and\ \bibinfo {author} {\bibfnamefont {T.~S.}\
  \bibnamefont {Koivisto}},\ }\bibfield  {title} {\bibinfo {title} {{General
  teleparallel quadratic gravity}},\ }\href
  {https://doi.org/10.1016/j.physletb.2020.135422} {\bibfield  {journal}
  {\bibinfo  {journal} {Phys. Lett. B}\ }\textbf {\bibinfo {volume} {805}},\
  \bibinfo {pages} {135422} (\bibinfo {year} {2020}{\natexlab{a}})},\ \Eprint
  {https://arxiv.org/abs/1909.09045} {arXiv:1909.09045 [gr-qc]} \BibitemShut
  {NoStop}%
\bibitem [{\citenamefont {Boehmer}\ and\ \citenamefont
  {Jensko}(2021)}]{Boehmer:2021aji}%
  \BibitemOpen
  \bibfield  {author} {\bibinfo {author} {\bibfnamefont {C.~G.}\ \bibnamefont
  {Boehmer}}\ and\ \bibinfo {author} {\bibfnamefont {E.}~\bibnamefont
  {Jensko}},\ }\bibfield  {title} {\bibinfo {title} {{Modified gravity: A
  unified approach}},\ }\href {https://doi.org/10.1103/PhysRevD.104.024010}
  {\bibfield  {journal} {\bibinfo  {journal} {Phys. Rev. D}\ }\textbf {\bibinfo
  {volume} {104}},\ \bibinfo {pages} {024010} (\bibinfo {year} {2021})},\
  \Eprint {https://arxiv.org/abs/2103.15906} {arXiv:2103.15906 [gr-qc]}
  \BibitemShut {NoStop}%
\bibitem [{\citenamefont {Beltr\'an~Jim\'enez}\ and\ \citenamefont
  {Koivisto}(2021)}]{BeltranJimenez:2021auj}%
  \BibitemOpen
  \bibfield  {author} {\bibinfo {author} {\bibfnamefont {J.}~\bibnamefont
  {Beltr\'an~Jim\'enez}}\ and\ \bibinfo {author} {\bibfnamefont {T.~S.}\
  \bibnamefont {Koivisto}},\ }\bibfield  {title} {\bibinfo {title} {{Accidental
  gauge symmetries of Minkowski spacetime in Teleparallel theories}},\ }\href
  {https://doi.org/10.3390/universe7050143} {\bibfield  {journal} {\bibinfo
  {journal} {Universe}\ }\textbf {\bibinfo {volume} {7}},\ \bibinfo {pages}
  {143} (\bibinfo {year} {2021})},\ \Eprint {https://arxiv.org/abs/2104.05566}
  {arXiv:2104.05566 [gr-qc]} \BibitemShut {NoStop}%
\bibitem [{\citenamefont {Beltr\'an~Jim\'enez}\ and\ \citenamefont
  {Koivisto}(2022)}]{BeltranJimenez:2021kpj}%
  \BibitemOpen
  \bibfield  {author} {\bibinfo {author} {\bibfnamefont {J.}~\bibnamefont
  {Beltr\'an~Jim\'enez}}\ and\ \bibinfo {author} {\bibfnamefont {T.~S.}\
  \bibnamefont {Koivisto}},\ }\bibfield  {title} {\bibinfo {title} {{Noether
  charges in the geometrical trinity of gravity}},\ }\href
  {https://doi.org/10.1103/PhysRevD.105.L021502} {\bibfield  {journal}
  {\bibinfo  {journal} {Phys. Rev. D}\ }\textbf {\bibinfo {volume} {105}},\
  \bibinfo {pages} {L021502} (\bibinfo {year} {2022})},\ \Eprint
  {https://arxiv.org/abs/2111.04716} {arXiv:2111.04716 [gr-qc]} \BibitemShut
  {NoStop}%
\bibitem [{\citenamefont {Hohmann}(2022)}]{Hohmann:2022mlc}%
  \BibitemOpen
  \bibfield  {author} {\bibinfo {author} {\bibfnamefont {M.}~\bibnamefont
  {Hohmann}},\ }\bibfield  {title} {\bibinfo {title} {{Teleparallel gravity}},\
  }\href@noop {} {\  (\bibinfo {year} {2022})},\ \Eprint
  {https://arxiv.org/abs/2207.06438} {arXiv:2207.06438 [gr-qc]} \BibitemShut
  {NoStop}%
\bibitem [{\citenamefont {Minkevich}\ and\ \citenamefont
  {Garkun}(1998)}]{Minkevich:1998cv}%
  \BibitemOpen
  \bibfield  {author} {\bibinfo {author} {\bibfnamefont {A.~V.}\ \bibnamefont
  {Minkevich}}\ and\ \bibinfo {author} {\bibfnamefont {A.~S.}\ \bibnamefont
  {Garkun}},\ }\bibfield  {title} {\bibinfo {title} {{Isotropic cosmology in
  metric - affine gauge theory of gravity}},\ }\href@noop {} {\  (\bibinfo
  {year} {1998})},\ \Eprint {https://arxiv.org/abs/gr-qc/9805007}
  {arXiv:gr-qc/9805007 [gr-qc]} \BibitemShut {NoStop}%
%%CITATION = GR-QC/9805007;%%
\bibitem [{\citenamefont {Hohmann}(2020)}]{Hohmann:2019fvf}%
  \BibitemOpen
  \bibfield  {author} {\bibinfo {author} {\bibfnamefont {M.}~\bibnamefont
  {Hohmann}},\ }\bibfield  {title} {\bibinfo {title} {{Metric-affine Geometries
  With Spherical Symmetry}},\ }\href {https://doi.org/10.3390/sym12030453}
  {\bibfield  {journal} {\bibinfo  {journal} {Symmetry}\ }\textbf {\bibinfo
  {volume} {12}},\ \bibinfo {pages} {453} (\bibinfo {year} {2020})},\ \Eprint
  {https://arxiv.org/abs/1912.12906} {arXiv:1912.12906 [math-ph]} \BibitemShut
  {NoStop}%
%%CITATION = ARXIV:1912.12906;%%
\bibitem [{\citenamefont {Hohmann}(2021{\natexlab{a}})}]{Hohmann:2021ast}%
  \BibitemOpen
  \bibfield  {author} {\bibinfo {author} {\bibfnamefont {M.}~\bibnamefont
  {Hohmann}},\ }\bibfield  {title} {\bibinfo {title} {{General covariant
  symmetric teleparallel cosmology}},\ }\href
  {https://doi.org/10.1103/PhysRevD.104.124077} {\bibfield  {journal} {\bibinfo
   {journal} {Phys. Rev. D}\ }\textbf {\bibinfo {volume} {104}},\ \bibinfo
  {pages} {124077} (\bibinfo {year} {2021}{\natexlab{a}})},\ \Eprint
  {https://arxiv.org/abs/2109.01525} {arXiv:2109.01525 [gr-qc]} \BibitemShut
  {NoStop}%
\bibitem [{\citenamefont {D'Ambrosio}\ \emph {et~al.}(2021)\citenamefont
  {D'Ambrosio}, \citenamefont {Heisenberg},\ and\ \citenamefont
  {Kuhn}}]{DAmbrosio:2021pnd}%
  \BibitemOpen
  \bibfield  {author} {\bibinfo {author} {\bibfnamefont {F.}~\bibnamefont
  {D'Ambrosio}}, \bibinfo {author} {\bibfnamefont {L.}~\bibnamefont
  {Heisenberg}},\ and\ \bibinfo {author} {\bibfnamefont {S.}~\bibnamefont
  {Kuhn}},\ }\bibfield  {title} {\bibinfo {title} {{Revisiting Cosmologies in
  Teleparallelism}},\ }\href@noop {} {\  (\bibinfo {year} {2021})},\ \Eprint
  {https://arxiv.org/abs/2109.04209} {arXiv:2109.04209 [gr-qc]} \BibitemShut
  {NoStop}%
\bibitem [{\citenamefont {Hohmann}(2021{\natexlab{b}})}]{Hohmann:2020zre}%
  \BibitemOpen
  \bibfield  {author} {\bibinfo {author} {\bibfnamefont {M.}~\bibnamefont
  {Hohmann}},\ }\bibfield  {title} {\bibinfo {title} {{Complete classification
  of cosmological teleparallel geometries}}\ }(\bibinfo {year} {2021})\ p.\
  \bibinfo {pages} {2140005},\ \Eprint {https://arxiv.org/abs/2008.12186}
  {arXiv:2008.12186 [gr-qc]} \BibitemShut {NoStop}%
%%CITATION = ARXIV:2008.12186;%%
\bibitem [{\citenamefont {Iosifidis}(2020)}]{Iosifidis:2020gth}%
  \BibitemOpen
  \bibfield  {author} {\bibinfo {author} {\bibfnamefont {D.}~\bibnamefont
  {Iosifidis}},\ }\bibfield  {title} {\bibinfo {title} {{Cosmological
  Hyperfluids, Torsion and Non-metricity}},\ }\href
  {https://doi.org/10.1140/epjc/s10052-020-08634-z} {\bibfield  {journal}
  {\bibinfo  {journal} {Eur. Phys. J.}\ }\textbf {\bibinfo {volume} {C80}},\
  \bibinfo {pages} {1042} (\bibinfo {year} {2020})},\ \Eprint
  {https://arxiv.org/abs/2003.07384} {arXiv:2003.07384 [gr-qc]} \BibitemShut
  {NoStop}%
%%CITATION = ARXIV:2003.07384;%%
\bibitem [{\citenamefont {Hohmann}(2021{\natexlab{c}})}]{Hohmann:2021fpr}%
  \BibitemOpen
  \bibfield  {author} {\bibinfo {author} {\bibfnamefont {M.}~\bibnamefont
  {Hohmann}},\ }\bibfield  {title} {\bibinfo {title} {{Variational Principles
  in Teleparallel Gravity Theories}},\ }\href
  {https://doi.org/10.3390/universe7050114} {\bibfield  {journal} {\bibinfo
  {journal} {Universe}\ }\textbf {\bibinfo {volume} {7}},\ \bibinfo {pages}
  {114} (\bibinfo {year} {2021}{\natexlab{c}})},\ \Eprint
  {https://arxiv.org/abs/2104.00536} {arXiv:2104.00536 [gr-qc]} \BibitemShut
  {NoStop}%
%%CITATION = ARXIV:2104.00536;%%
\bibitem [{\citenamefont {Yano}(1957)}]{Yano:1957lda}%
  \BibitemOpen
  \bibfield  {author} {\bibinfo {author} {\bibfnamefont {K.}~\bibnamefont
  {Yano}},\ }\href@noop {} {\emph {\bibinfo {title} {The Theory of Lie
  Derivatives and its Applications}}}\ (\bibinfo  {publisher} {North-Holland},\
  \bibinfo {address} {Amsterdam},\ \bibinfo {year} {1957})\BibitemShut
  {NoStop}%
\bibitem [{\citenamefont {Bahamonde}\ \emph {et~al.}(2022)\citenamefont
  {Bahamonde}, \citenamefont {Dialektopoulos}, \citenamefont {Hohmann},
  \citenamefont {Levi~Said}, \citenamefont {Pfeifer},\ and\ \citenamefont
  {Saridakis}}]{Bahamonde:2022ohm}%
  \BibitemOpen
  \bibfield  {author} {\bibinfo {author} {\bibfnamefont {S.}~\bibnamefont
  {Bahamonde}}, \bibinfo {author} {\bibfnamefont {K.~F.}\ \bibnamefont
  {Dialektopoulos}}, \bibinfo {author} {\bibfnamefont {M.}~\bibnamefont
  {Hohmann}}, \bibinfo {author} {\bibfnamefont {J.}~\bibnamefont {Levi~Said}},
  \bibinfo {author} {\bibfnamefont {C.}~\bibnamefont {Pfeifer}},\ and\ \bibinfo
  {author} {\bibfnamefont {E.~N.}\ \bibnamefont {Saridakis}},\ }\bibfield
  {title} {\bibinfo {title} {{Perturbations in Non-Flat Cosmology for $f(T)$
  gravity}},\ }\href@noop {} {\  (\bibinfo {year} {2022})},\ \Eprint
  {https://arxiv.org/abs/2203.00619} {arXiv:2203.00619 [gr-qc]} \BibitemShut
  {NoStop}%
\bibitem [{\citenamefont {Hohmann}(2018)}]{Hohmann:2018ijr}%
  \BibitemOpen
  \bibfield  {author} {\bibinfo {author} {\bibfnamefont {M.}~\bibnamefont
  {Hohmann}},\ }\bibfield  {title} {\bibinfo {title} {{Scalar-torsion theories
  of gravity III: analogue of scalar-tensor gravity and conformal
  invariants}},\ }\href {https://doi.org/10.1103/PhysRevD.98.064004} {\bibfield
   {journal} {\bibinfo  {journal} {Phys. Rev. D}\ }\textbf {\bibinfo {volume}
  {98}},\ \bibinfo {pages} {064004} (\bibinfo {year} {2018})},\ \Eprint
  {https://arxiv.org/abs/1801.06531} {arXiv:1801.06531 [gr-qc]} \BibitemShut
  {NoStop}%
\bibitem [{\citenamefont {Golovnev}\ and\ \citenamefont
  {Guzm\'an}(2021)}]{Golovnev:2020zpv}%
  \BibitemOpen
  \bibfield  {author} {\bibinfo {author} {\bibfnamefont {A.}~\bibnamefont
  {Golovnev}}\ and\ \bibinfo {author} {\bibfnamefont {M.-J.}\ \bibnamefont
  {Guzm\'an}},\ }\bibfield  {title} {\bibinfo {title} {{Foundational issues in
  f(T) gravity theory}},\ }\href {https://doi.org/10.1142/S0219887821400077}
  {\bibfield  {journal} {\bibinfo  {journal} {Int. J. Geom. Meth. Mod. Phys.}\
  }\textbf {\bibinfo {volume} {18}},\ \bibinfo {pages} {2140007} (\bibinfo
  {year} {2021})},\ \Eprint {https://arxiv.org/abs/2012.14408}
  {arXiv:2012.14408 [gr-qc]} \BibitemShut {NoStop}%
\bibitem [{\citenamefont {Golovnev}\ and\ \citenamefont
  {Koivisto}(2018)}]{Golovnev:2018wbh}%
  \BibitemOpen
  \bibfield  {author} {\bibinfo {author} {\bibfnamefont {A.}~\bibnamefont
  {Golovnev}}\ and\ \bibinfo {author} {\bibfnamefont {T.}~\bibnamefont
  {Koivisto}},\ }\bibfield  {title} {\bibinfo {title} {{Cosmological
  perturbations in modified teleparallel gravity models}},\ }\href
  {https://doi.org/10.1088/1475-7516/2018/11/012} {\bibfield  {journal}
  {\bibinfo  {journal} {JCAP}\ }\textbf {\bibinfo {volume} {11}},\ \bibinfo
  {pages} {012}},\ \Eprint {https://arxiv.org/abs/1808.05565} {arXiv:1808.05565
  [gr-qc]} \BibitemShut {NoStop}%
\bibitem [{\citenamefont {Golovnev}\ and\ \citenamefont
  {Guzman}(2021)}]{Golovnev:2020nln}%
  \BibitemOpen
  \bibfield  {author} {\bibinfo {author} {\bibfnamefont {A.}~\bibnamefont
  {Golovnev}}\ and\ \bibinfo {author} {\bibfnamefont {M.-J.}\ \bibnamefont
  {Guzman}},\ }\bibfield  {title} {\bibinfo {title} {{Nontrivial Minkowski
  backgrounds in $f(T)$ gravity}},\ }\href
  {https://doi.org/10.1103/PhysRevD.103.044009} {\bibfield  {journal} {\bibinfo
   {journal} {Phys. Rev. D}\ }\textbf {\bibinfo {volume} {103}},\ \bibinfo
  {pages} {044009} (\bibinfo {year} {2021})},\ \Eprint
  {https://arxiv.org/abs/2012.00696} {arXiv:2012.00696 [gr-qc]} \BibitemShut
  {NoStop}%
\bibitem [{\citenamefont {Jim\'enez}\ \emph {et~al.}(2021)\citenamefont
  {Jim\'enez}, \citenamefont {Golovnev}, \citenamefont {Koivisto},\ and\
  \citenamefont {Veerm\"ae}}]{Jimenez:2020ofm}%
  \BibitemOpen
  \bibfield  {author} {\bibinfo {author} {\bibfnamefont {J.~B.}\ \bibnamefont
  {Jim\'enez}}, \bibinfo {author} {\bibfnamefont {A.}~\bibnamefont {Golovnev}},
  \bibinfo {author} {\bibfnamefont {T.}~\bibnamefont {Koivisto}},\ and\
  \bibinfo {author} {\bibfnamefont {H.}~\bibnamefont {Veerm\"ae}},\ }\bibfield
  {title} {\bibinfo {title} {{Minkowski space in $f(T)$ gravity}},\ }\href
  {https://doi.org/10.1103/PhysRevD.103.024054} {\bibfield  {journal} {\bibinfo
   {journal} {Phys. Rev. D}\ }\textbf {\bibinfo {volume} {103}},\ \bibinfo
  {pages} {024054} (\bibinfo {year} {2021})},\ \Eprint
  {https://arxiv.org/abs/2004.07536} {arXiv:2004.07536 [gr-qc]} \BibitemShut
  {NoStop}%
\bibitem [{\citenamefont {Blagojevi\'c}\ and\ \citenamefont
  {Nester}(2020)}]{Blagojevic:2020dyq}%
  \BibitemOpen
  \bibfield  {author} {\bibinfo {author} {\bibfnamefont {M.}~\bibnamefont
  {Blagojevi\'c}}\ and\ \bibinfo {author} {\bibfnamefont {J.~M.}\ \bibnamefont
  {Nester}},\ }\bibfield  {title} {\bibinfo {title} {{Local symmetries and
  physical degrees of freedom in $f(T)$ gravity: a Dirac Hamiltonian constraint
  analysis}},\ }\href {https://doi.org/10.1103/PhysRevD.102.064025} {\bibfield
  {journal} {\bibinfo  {journal} {Phys. Rev. D}\ }\textbf {\bibinfo {volume}
  {102}},\ \bibinfo {pages} {064025} (\bibinfo {year} {2020})},\ \Eprint
  {https://arxiv.org/abs/2006.15303} {arXiv:2006.15303 [gr-qc]} \BibitemShut
  {NoStop}%
\bibitem [{\citenamefont {Guzm\'an}\ and\ \citenamefont
  {Ferraro}(2020)}]{Guzman:2019oth}%
  \BibitemOpen
  \bibfield  {author} {\bibinfo {author} {\bibfnamefont {M.~J.}\ \bibnamefont
  {Guzm\'an}}\ and\ \bibinfo {author} {\bibfnamefont {R.}~\bibnamefont
  {Ferraro}},\ }\bibfield  {title} {\bibinfo {title} {{Degrees of freedom and
  Hamiltonian formalism for $f(T)$ gravity}},\ }\href
  {https://doi.org/10.1142/S0217751X20400229} {\bibfield  {journal} {\bibinfo
  {journal} {Int. J. Mod. Phys. A}\ }\textbf {\bibinfo {volume} {35}},\
  \bibinfo {pages} {2040022} (\bibinfo {year} {2020})},\ \Eprint
  {https://arxiv.org/abs/1910.03100} {arXiv:1910.03100 [gr-qc]} \BibitemShut
  {NoStop}%
\bibitem [{\citenamefont {Beltr\'an~Jim\'enez}\ \emph
  {et~al.}(2020{\natexlab{b}})\citenamefont {Beltr\'an~Jim\'enez},
  \citenamefont {Heisenberg}, \citenamefont {Koivisto},\ and\ \citenamefont
  {Pekar}}]{BeltranJimenez:2019tme}%
  \BibitemOpen
  \bibfield  {author} {\bibinfo {author} {\bibfnamefont {J.}~\bibnamefont
  {Beltr\'an~Jim\'enez}}, \bibinfo {author} {\bibfnamefont {L.}~\bibnamefont
  {Heisenberg}}, \bibinfo {author} {\bibfnamefont {T.~S.}\ \bibnamefont
  {Koivisto}},\ and\ \bibinfo {author} {\bibfnamefont {S.}~\bibnamefont
  {Pekar}},\ }\bibfield  {title} {\bibinfo {title} {{Cosmology in $f(Q)$
  geometry}},\ }\href {https://doi.org/10.1103/PhysRevD.101.103507} {\bibfield
  {journal} {\bibinfo  {journal} {Phys. Rev. D}\ }\textbf {\bibinfo {volume}
  {101}},\ \bibinfo {pages} {103507} (\bibinfo {year} {2020}{\natexlab{b}})},\
  \Eprint {https://arxiv.org/abs/1906.10027} {arXiv:1906.10027 [gr-qc]}
  \BibitemShut {NoStop}%
\end{thebibliography}%

\end{document}